\documentclass[prb,aps,superscriptaddress,showpacs,showkeys]{revtex4} 
\usepackage{graphicx}
\usepackage{amsmath}
\usepackage{epsfig}
\begin{document}

\title{Hybrid path integral Monte Carlo simulation of rigid diatomic
  molecules: effect of quantized rotations on the selectivity of hydrogen isotopes in carbon nanotubes}

\author{Giovanni Garberoglio} 
\altaffiliation{Currently staying at Dipartimento di Fisica, Universit\`a
  degli Studi di Trento, Italy}
\email{garberog@science.unitn.it} 
\affiliation{Department of Chemical and Petroleum Engineering, University of
Pittsburgh, Pittsburgh, PA 15261, USA} 
\affiliation{National Energy Technology Laboratory, Pittsburgh, PA, 15236
  USA} 

\author{J. Karl Johnson} 
\affiliation{Department of Chemical and Petroleum Engineering, University of
  Pittsburgh, Pittsburgh, PA 15261, USA} 
\affiliation{National Energy Technology Laboratory, Pittsburgh, PA, 15236
  USA} 

\date{\today}

\begin{abstract}
We present a multiple time step algorithm for hybrid path integral Monte Carlo
simulations involving rigid linear rotors.  We show how to calculate the
quantum torques needed in the simulation from the rotational density matrix,
for which we develop an approximate expression suitable in the case of
heteronuclear molecules.

We use this method to study the effect of rotational quantization on the
quantum sieving properties of carbon nanotubes, with particular emphasis to
the {\em para}-T${}_2$/{\em para}-H${}_2$ selectivity at 20~K. We show how to
treat classically only some of the degrees of freedom of the hydrogen molecule
and we find that in the limit of zero pressure the quantized nature of the
rotational degrees of freedom greatly influence the selectivity, especially in
the case of the (3,6) nanotube, which is the narrowest tube that we have
studied.

We also use path integral Monte Carlo simulations to calculate adsorption
isotherms of different hydrogen isotopes in the interior of carbon nanotubes
and the corresponding selectivity at finite pressures. It is found that the
selectivity increases with respect to the zero pressure value and tends to a
constant value at saturation. We use a simplified effective harmonic
oscillator model to discuss the origin of this behavior.
\end{abstract}


\maketitle

\section{Introduction}

Carbon nanotubes have received a lot of attention in the past years in many
fields of science and engineering, ranging from metallic properties, to
adsorption of light gases.

The availability of these potentially narrow (subnanometer) channels has
raised the interest about the possibility of using them as isotope sieves.
The presence of quantum effects on adsorption has been known for quite a long
time~\cite{KatorskiW64,Moiseyev75} and Krylov {\em et al.}~\cite{krylov} were
the first to show, in the framework of a simple rigid wall model, the
possibility of separating helium and hydrogen isotopes by using the different
zero-point energy of the two confined species.

The idea was developed by Johnson and
collaborators~\cite{wang,challa_zerop,challa}, with particular emphasis about
the calculation of the selectivity of T$_2$/H$_2$ mixtures in the interior
channels of carbon nanotubes. These authors used an interaction potential
model which neglects the molecular structure of hydrogen and developed a
method - known as the simple theory - to calculate the selectivity in the
limit of zero pressure (i.e. neglecting hydrogen-hydrogen interactions) by
using the single particle energy levels in the confined system.~\cite{wang,
  challa_zerop}

They conclude that the (3,6) nanotube would be able to show a selectivity of
the order of 10$^5$ at zero pressure and 20~K and calculated the expected
selectivities for a wide set of nanotubes, thus displaying the dependence of
this quantity on the radius of the tubes.
They later extended their calculations, using path integral Monte Carlo
methods, to finite pressures.~\cite{challa}

Hathorn {\em et al.}~\cite{hathorn} were the first to address the effect of
the rotational degrees of freedom of hydrogen on the selectivity. Assuming a
decoupling of the rotational and translational motions they showed that in
narrow tubes one can expect an increase of the selectivity in narrow tubes of
a factor 100 at 20~K, when compared with models that approximate hydrogen as a
sphere.

These results have been confirmed by Trasca {\em et al.}~\cite{trasca}, who
calculated the D$_2$/H$_2$ selectivity in the interstitial channels and groove
sites of various carbon nanotubes bundles.

Lu {\em et al.}~\cite{goldfield} have also calculated, in the framework
of the simple theory, the energy levels and selectivity of molecular hydrogen
in carbon nanotubes, by numerical diagonalization of the Schr\"odinger
equation of a confined rotor, finding a value of the total selectivity for
T$_2$/H$_2$ mixtures at zero pressure and 20~K of the order of
100 in the (3,6) tube, a result far lower than the ones already published in
the literature.
This result was later attributed to the use of an unphysical potential for the
hydrogen-carbon interaction. Calculations with more realistic
potentials~\cite{Garberoglio2006,goldfield2006} have confirmed the expectation
of high selectivity.

Two problems seem still without a definite solution: the first is the
effectiveness of model potentials in describing the carbon-hydrogen
interactions. Different authors have used different potentials, and their
predicted values of the selectivity scatter in an enormous range, from 10$^2$
to 10$^7$.
The second point, quite related to the first, is the actual effect of the
rotational degrees of freedom on the selectivity.

We have addressed both these issues in a recent paper~\cite{Garberoglio2006},
where, in the framework of the simple theory, we have calculated the
dependence of the zero pressure selectivity on the hydrogen-carbon interaction
potential and we have developed an approximate method to evaluate the
contribution of rotational-translational coupling to the selectivity itself.

We have demonstrated that the zero pressure selectivity is very much dependent
on even small changes in the interaction potential, especially in the case of
narrow tubes. Moreover the presence of a steep confining potential results in
a strong translational-rotational coupling, and the overall selectivity cannot
be calculated neither by assuming independence between those two degrees of
freedom nor by assuming a spherically symmetric model for the hydrogen
molecule. 
These results have been validated by the exact diagonalization of the
hydrogen-carbon Hamiltonian reported in a recent paper by Lu {\em et
  al.}~\cite{goldfield2006}

In this paper we address the issue of the contribution of rotational degrees
of freedom in more detail, using the path integral Monte Carlo method to treat
the quantized rotational degrees of freedom in an exact way.  After discussing
how to perform an efficient path integral simulation of rotors using the
hybrid Monte Carlo (HMC) technique, we develop a method for the classical
treatment of the rotational degrees of freedom only and show that it gives a
much lower zero pressure selectivity than the full quantum model. We also
extend the calculations at finite pressures and discuss how the selectivity
changes in this regime.

This paper is organized as follows. In Sec.~\ref{sec:HMC} we describe the
hybrid Monte Carlo method that we have used in this work. We further address
the issue on how the zero pressure selectivity can be calculated from path
integral simulations, with particular emphasis to the influence of the
quantized rotational degrees of freedom. We develop a formalism that enables
one to simulate classical rotations together with quantized center of mass
degrees of freedom using a path integral approach and use this method to
investigate the effect of the quantized rotational degrees of freedom on the
selectivity..

In Sec.~\ref{sec:results} we present and discuss our results.
Some technical derivations are presented in the appendices.

\section{The hybrid path integral Monte Carlo method}
\label{sec:HMC}

\subsection{The path integral formulation of statistical mechanics}

The quantum mechanical expression for the partition function of a system of
$N$ rigid linear rotors is
\begin{equation}
  Q = \int d^{3N}X_1 d^{2N}\Omega_1 ~ \langle X_1 \Omega_1 | \exp[-\beta
  \hat{H}] | X_1 \Omega_1 \rangle
\label{eq:Q}
\end{equation}
where we denote with $X$ a vector with the $3N$ center of mass coordinates of
the rotors, and with $\Omega$ the set of the $2N$ angles describing their
orientations. A subscript $1$ has been introduced for later convenience.
The Hamiltonian $\hat{H}$ is given by
\begin{equation}
  \hat{H} = \sum_{i=1}^{N} - \frac{\hbar^2}{2M} \nabla_i^2 + 
            \sum_{i=1}^{N} \frac{\hat{L}_i^2}{2 I} +
	    \sum_{i<j} v({\hat x}_i {\hat \omega}_i ; {\hat x}_j {\hat \omega}_j) +
	    \sum_{i=1}^N v_{\rm ext}({\hat x}_i {\hat \omega}_i)
\label{eq:H}
\end{equation}
and is a function of the mass $M$, the angular momentum ${\hat L}$ and the
inertia moment $I$ of the molecules. We have introduced a fluid-fluid pair
interaction potential $v$ and a solid-fluid interaction potential $v_{\rm
ext}$. The quantities ${\hat x}_i$ and ${\hat \omega}_i$ denote the operators
for the center of mass position and the orientation of molecule $i$,
respectively. The Hamiltonian is, as usual, the sum of the translational
(center of mass) kinetic energy $\hat{T}$, the rotational kinetic energy
$\hat{K}$, the fluid-fluid potential energy $\hat{V}$ and the solid-fluid
potential energy $\hat{V}_{\rm ext}$.

The quantum partition function of Eq.~(\ref{eq:Q}) can be rewritten using
repeatedly the Trotter identity
\begin{equation}
  \exp[A+B] = \lim_{P \rightarrow \infty} 
  \left( \exp[A/P] ~ \exp[B/P] \right)^P
\end{equation}
valid for generally non-commuting operators $A$ and $B$ and further
approximated by assuming a large but finite Trotter number $P$, obtaining
\begin{equation}
  Q \simeq \int d^{3N}X_1 d^{2N}\Omega_1 ~ \langle X_1 \Omega_1 |
  \prod_{j=1}^{P} \exp(-\beta \hat{T}/P) \exp(-\beta \hat{K}/P) 
  \exp[-\beta (\hat{V} + \hat{V}_{\rm ext})/P]
  | X_1 \Omega_1 \rangle
\label{eq:PIQ}
\end{equation}

We can now introduce $P-1$ completeness relations of the form
\begin{equation}
{\mathbf 1} = \int d^{3N}X_i d^{2N}\Omega_i ~ 
    | X_i \Omega_i \rangle \langle X_i \Omega_i |
\end{equation}
between the factors in Eq.~(\ref{eq:PIQ}) and write the partition function
as
\begin{eqnarray}
  Q &\simeq& \
  \int d^{3N}X_1 d^{2N}\Omega_1 \ldots d^{3N}X_P d^{2N}\Omega_P \nonumber \\
  & & \langle X_1 \Omega_1 |
  \exp(-\beta \hat{T}/P) \exp(-\beta \hat{K}/P) 
  \exp[-\beta (\hat{V} + \hat{V}_{\rm ext})/P]
  | X_2 \Omega_2 \rangle \nonumber \\
  & & \langle X_2 \Omega_2 |
  \exp(-\beta \hat{T}/P) \exp(-\beta \hat{K}/P) 
  \exp[-\beta (\hat{V} + \hat{V}_{\rm ext})/P] 
  | X_3 \Omega_3 \rangle \nonumber \\
  & & \ldots \nonumber \\
  & & \langle X_P \Omega_P |
  \exp(-\beta \hat{T}/P) \exp(-\beta \hat{K}/P) 
  \exp[-\beta (\hat{V} + \hat{V}_{\rm ext})/P]
  | X_1 \Omega_1 \rangle \nonumber \\
  &=&   \int d^{3N}X_1 d^{2N}\Omega_1 \ldots d^{3N}X_P d^{2N}\Omega_P
  \nonumber \\
  &&  \prod_{t=1}^P \langle X_t \Omega_t |
  \exp(-\beta \hat{T}/P) \exp(-\beta \hat{K}/P) 
  \exp[-\beta (\hat{V} + \hat{V}_{\rm ext})/P]
  | X_{t+1} \Omega_{t+1} \rangle
  \label{eq:PIQ_2}
\end{eqnarray}
where we have denoted $X_{P+1} = X_1$ and $\Omega_{P+1} = \Omega_1$.
Each of the matrix elements appearing in the previous equation can be
written as
\begin{eqnarray}
  \langle X_i \Omega_i |
  \exp(-\beta \hat{T}/P) \exp(-\beta \hat{K}/P) 
  \exp[-\beta (\hat{V} + \hat{V}_{\rm ext})/P]
  | X_{i+1} \Omega_{i+1} \rangle &=& \nonumber \\ 
  \langle X_i | \exp(-\beta \hat{T}/P) | X_{i+1} \rangle ~
  \langle \Omega_i | \exp(-\beta \hat{K}/P) | \Omega_{i+1} \rangle ~
  \exp[-\beta (V(X_{i+1} \Omega_{i+1}) + V_{\rm ext}(X_{i+1}
  \Omega_{i+1}))/P] 
\label{eq:factor}
\end{eqnarray}
A straightforward calculation shows that the expectation value of the
translational kinetic energy Boltzmann factor assumes the form~\cite{binder}
\begin{equation}
  \langle X_i | \exp(-\beta \hat{T}/P) | X_{i+1} \rangle = 
   a
  \exp(-\beta \kappa |X_i - X_{i+1}|^2/2)
\label{eq:tke}
\end{equation}
where the amplitude $a$ and the ``spring constant'' $\kappa$ are given by
\begin{eqnarray}
  a &=& \left( \frac{M k_B T P}{2 \pi \hbar^2} \right)^{3/2}
  \label{eq:a} \\
  \kappa &=& \frac{M P (k_B T)^2}{\hbar^2}
\label{eq:kappa}
\end{eqnarray}
and the expectation value of the rotational kinetic energy Boltzmann factor
becomes~\cite{rotors}
\begin{eqnarray}
  \langle \Omega_i | \exp(-\beta \hat{K}/P) | \Omega_{i+1} \rangle &=& 
  \sum_{n=1}^N
  \sum_{J=0}^{\infty} \frac{2J+1}{4 \pi} P_J(\cos\theta^n_{i,i+1})
  \exp[-\beta J(J+1) B/P] \nonumber \\
  &\equiv& \sum_{n=1}^N \Xi(\theta^n_{i,i+1})
\label{eq:rot_dm}
\end{eqnarray}
where $P_J(\cdot)$ is a Legendre polynomial, $\theta^n_{i,i+1}$ is the
angle between the directions $\omega_i$ and $\omega_{i+1}$ relative to
molecule $n$ 
and $B = \hbar^2/(2 I)$ is the rotational constant of the rotor.
In the case of homonuclear molecules the indistinguishability of the nuclei
imposes some restrictions on the sum in Eq.~(\ref{eq:rot_dm}) according to the
spin states of the nuclei: for {\em para}-H$_2$, {\em ortho}-D$_2$
and {\em para}-T$_2$, as in this work, the summation on the angular momenta
$J$ in Eq.~(\ref{eq:rot_dm}) is limited to the even numbers
only~\cite{rotors} and results in a positive definite density matrix, which
can be directly used in the Monte Carlo simulations.

In the other rotational states (i.e. {\em ortho}-H$_2$ and T$_2$ and 
{\em para}-D$_2$) the sum is restricted to the odd angular momentum states,
resulting in a density matrix which is not positive definite. As a
consequence, more care has to be taken in performing a Monte Carlo
simulation in this case~\cite{rotors}.

The net effect of these algebraic manipulations is that we have been able to
rewrite the original quantum partition function of an $N$ particle system,
Eq.~(\ref{eq:Q}), as a classical partition function of a $NP$ particle
systems.
The $NP$ particle of the classical equivalent are naturally divided in $P$
subsets (also known as time slices) of $N$ particles each. Each particle in a
time slice interacts with all the other particles in the same time slice via
the original intermolecular and intramolecular potential divided by a factor
of $P$ (see Eq.~(\ref{eq:factor})). Quantum mechanical effects taken into
account by the interaction of each particle with the corresponding copy on the
previous and following time slice: the center of mass coordinates are bound by
the harmonic potential of Eq.~(\ref{eq:tke}) and the orientations give rise to
the inter-slice rotational partition function of Eq.~(\ref{eq:rot_dm}). The
resulting system is then equivalent to a classical collection of $N$ ring
polymers, each having $P$ beads. The $i$-th particle on a given polymer
interacts only with the corresponding particle on the other polymers via the
original intermolecular potential (rescaled by a factor of $P$). The
interaction between the beads of a given ring polymer are described by an
harmonic interaction on the translational coordinates with the two adjacent
beads (see Eq.~(\ref{eq:tke})) and an interaction between the orientational
degrees of freedom of adjacent beads whose density matrix is given by
Eq.~(\ref{eq:rot_dm}).

\subsection{Observables and estimators}

In the framework of the path integral approximation to the quantum partition
function the estimators for the average values of interest can usually be
obtained by their thermodynamic definitions.

The estimator for the translational kinetic energy is then~\cite{binder}
\begin{equation}
  T^{\rm est} = \frac{3}{2} k_B T P - U_{\rm quant}
  \label{eq:tke_est}
\end{equation}
where $U_{\rm quant} = (\kappa /2) \sum_{t=1}^P |X_t - X_{t+1}|^2$ is the
value of the quantum spring potential energy, and the one for the rotational
kinetic energy is~\cite{rotors}
\begin{eqnarray}
  K^{\rm est} &=& \frac{1}{P} \sum_{i=1}^P K^{\rm est}_i \\
  K^{\rm est}_i &=& \frac{B}{4 \pi \Xi} \sum_{J} J (J+1) (2J+1) ~
  P_J(\cos\theta_{i ~ i+1}) \exp\left( - \frac{\beta}{P} B J (J+1) \right)
  \label{eq:qke_est}
\end{eqnarray}

We can derive an expression for the calculation of the particle density in a
path integral simulation. Denoting $\hat{x}^{(t)}_i$ as the position
operator of particle $i$ in time slice $t$, one has
\begin{eqnarray}
  \rho(r) &=& \langle \sum_{n=1}^N \frac{1}{N} \delta(r - \hat{x}_n)
  \rangle \\
  &=& \frac{1}{Q} \int d^{3N}X_1 d^{2N} \Omega_1 ~ 
  \langle X_1 \Omega_1| \sum_{n=1}^N \frac{1}{N} \delta(r - \hat{x}^{(1)}_n)
  \exp[-\beta \hat{H}]| X_1 \Omega_1 \rangle \\
  &=& \frac{1}{Q} \int d^{3N}X_1 d^{2N} \Omega_1 ~ 
  \sum_{n=1}^N \frac{1}{N} \delta(r - x^{(1)}_n) \times \nonumber \\
  & &
  \langle X_1 \Omega_1 | 
  \prod_{j=1}^{P} \exp(-\frac{\beta}{P} \hat{T}) \exp(-\frac{\beta}{P}
  \hat{K})  \exp[-\frac{\beta}{P} (\hat{V} + \hat{V}_{\rm ext})]
  | X_1 \Omega_1 \rangle
  \label{eq:quasi_d}
\end{eqnarray}
introducing now $P-1$ completeness relations, analogously to the passage
leading to Eq.~(\ref{eq:PIQ_2}) and noting that the integrand can be rewritten
by relabeling $X_1 \leftrightarrow X_j$, one obtains from
Eq.(\ref{eq:quasi_d})
\begin{eqnarray}
  \rho(r) &=& \frac{1}{Q} \int \prod_{p=1}^P d^{3N}X_p d^{2N}\Omega_p 
  \left( \frac{1}{NP} \sum_{t=1}^P \sum_{n=1}^N \delta(r - x^{(t)}_n) 
  \right) \times \nonumber \\
  &&\prod_{p=1}^{P} 
  \langle X_p \Omega_p |  
  \exp(-\frac{\beta}{P} \hat{T}) \exp(-\frac{\beta}{P}
  \hat{K})  \exp[-\frac{\beta}{P} (\hat{V} + \hat{V}_{\rm ext})]
  | X_{p+1} \Omega_{p+1} \rangle
  \label{eq:density}
\end{eqnarray}
so that the density at a point $r$ is equivalent to the probability of finding
the center of mass of a bead at the same point.

\subsection{The hybrid Monte Carlo method}

Using the path integral formulation, a quantum partition function can be
rewritten as a classical partition function of a system with a larger number
of particles, so that classical Monte Carlo methods can then be used to
calculate thermodynamic properties. 
Since we expect to work at conditions where quantum mechanics is not a small
correction, it will be necessary to use large values of the Trotter number
$P$. A simple Metropolis method for the sampling of the translational and
rotational phase space, such as the one discussed in
Ref.~\onlinecite{rot_order}, 
will be affected by slow convergence. We have then decided to use the hybrid
Monte Carlo method,~\cite{hmc,hmc2} that consists in choosing a new candidate
configuration by performing a molecular dynamics (MD) move with a large
timestep; the resulting configuration is then accepted or rejected using a
standard Metropolis condition on the difference in the {\em total} energy
(which is not conserved if a large enough timestep is taken).

In order to perform an MD move one has to know the forces and the torques
acting on each of the rotors. The potential energy between the molecules on the
same slice is given by the rescaled original potential, and the quantum
mechanical effects on the translational degrees of freedom are described by a
simple harmonic potential between adjacent slices (see
Eq.~(\ref{eq:tke})). The only unknown is the quantum torque between the
molecules in adjacent slices: we have decided to calculate it numerically,
starting from the expression of the density matrix in Eq.~(\ref{eq:rot_dm})
(see Appendix~\ref{sec:app_rot} for an analytic limit in the case of
heteronuclear molecules). Since Eq.~(\ref{eq:rot_dm}) represents a Boltzmann
factor it can generally be written as $B_{\rm rot} = C \exp[-\beta U_{\rm
rot}(\theta)]$ where $C$ is an unknown constant and $U_{\rm rot}(\theta)$ is
the quantum rotational potential energy between two adjacent rotors, whose
orientations form an angle $\theta$ with one another; we show in the appendix
that for heteronuclear molecules in the high $P$ limit $U_{\rm rot}$ is given
to a good approximation by the harmonic expression $U_{\rm rot}(\theta) = K
\theta^2/2$ with $K$ given in Eq.~(\ref{eq:K}). In the
general case the modulus of the quantum torque can be written as
\begin{equation}
N_{\rm quant}(\theta) = - \frac{d U_{\rm rot}(\theta)}{d \theta} =
\frac{k_B T}{\Xi} \sin\theta \frac{d \Xi (\theta)}{d \cos\theta}
\label{eq:qtorque}
\end{equation}
The direction of the torque is obviously orthogonal to the plane generated by
the two orientations of the interacting molecules, and such that it tends to
close the angle between the two molecules. The derivative in
Eq.~(\ref{eq:qtorque}) could be evaluated either numerically or using the
identity
$$
\frac{d P_l(x)}{d x} = \frac{l x P_l(x) - l P_{l-1}(x)}{x^2-1}
$$

\subsection{The multiple time step method}

Inspection of Eqs.~(\ref{eq:factor}) and (\ref{eq:kappa}) 
shows that the intermolecular forces scale like the inverse of the Trotter
number $P$, whereas the quantum forces (and, possibly, the torques) are
proportional to it. In order to efficiently sample phase space using an
hybrid Monte Carlo method it is necessary that all the degrees of freedom
contribute uniformly to the non-conservation of energy when a MD time step is
performed. Since the intermolecular forces become weaker for large Trotter
number 
while the quantum forces become stronger,
we have decided to use a
multiple time step method to perform the MD evolution.~\cite{multiple-ts} We
divide the forces into ``long range'' (the intermolecular forces, in our case)
and ``short range'' (the quantum forces and torques). Each of the long range
dynamical steps $\Delta t$ is divided into $n$ smaller time steps $\delta t$
where only the short range forces are evaluated as the system is propagated.

In our case we do not know the typical time scale of the quantum rotation, so
we have decided to use three nested loops: we use a long time step $\Delta t$
to propagate the system according to the intermolecular forces, an
intermediate time step $\delta t$ to propagate the quantum spring forces on
the translational degrees of freedom and a rotational time step $\delta \tau$
to propagate the quantum torques.

We also need, for the hybrid method to work, a reversible algorithm to
integrate all the degrees of freedom. It is well known that the velocity form
of the Verlet algorithm possess such a feature and can be used in multiple
time step methods.~\cite{multiple-ts} Instead of using algorithms already
developed to treat the general motion of rigid rotors in a multiple time step
framework~\cite{rot-ts} we have developed a velocity Verlet like integrator
for the rotational motion of a rigid linear rotor that can be easily
integrated in the velocity Verlet evolution of the center of mass
coordinates. Details of its derivation are given in
Appendix~\ref{sec:appendix}.

Denoting by $x$ and $v$ the translational positions and velocities, and by $e$
and $\varpi$ the direction of the molecular axes and the molecular angular
velocities, the multiple time step method is then implemented as follows:

\begin{tabbing}
 \= Calculate $F_{long}$ and $N_{long}$  (intermolecular forces)  \= \=\\
 \> $v \rightarrow v + \Delta t F_{long} /(2M)$           \= \=\\
 \> $\varpi \rightarrow \varpi + \Delta t N_{long} /(2I)$ \= \=\\
 \> \> Calculate $F_{short}$ (quantum spring)                 \=\\
 \> \> $v \rightarrow v + \delta t F_{short} / (2M)$          \=\\
 \> \> \> Calculate $N_{short}$ (quantum torque)                \\
 \> \> \> $\varpi \rightarrow \varpi + \delta\tau N_{long} /(2I) $ \\
 \> \> \> $e \rightarrow e  + \delta \tau ~ \varpi \times e - (\delta \tau)^2
 \varpi^2 /2$\\
 \> \> \> normalize $e$ \\

 \> \> \> Calculate $N_{short}$ (quantum torque)                \\
 \> \> \> $\varpi \rightarrow \varpi + \delta\tau N_{long} /(2I) $ \\
 \> \> \> $x \rightarrow x + v \delta\tau $ \\

 \> \> Calculate $F_{short}$ \= \\
 \> \> $v \rightarrow v + \delta t F_{short} / (2M)$  \= \\
 \> Calculate $F_{long}$ and $N_{long}$ \= \= \\
 \> $v \rightarrow v + \Delta t F_{long} /(2M)$           \= \=\\
 \> $\varpi \rightarrow \varpi + \Delta t N_{long} /(2I)$ \= \=\\
 \> Calculate the final translational and rotational kinetic energy \= \= \\
\end{tabbing}

In order to fix the integer values of the ratios $\delta t / \delta \tau$
and $\Delta t / \delta t$ we have proceeded as follows. For the first ratio we
have performed some ideal gas simulations, where the translational and
rotational motions are decoupled and can be simulated by taking into account
translations and rotations separately. We have adjusted the rotational and
translational ideal gas time steps in order to have a 50\% 
acceptance ratio and we have then fixed $\delta t/ \delta \tau = (\delta t /
\delta \tau)_{\rm ideal\ gas} \simeq 3$.
 
The ratio $\Delta t/ \delta t$ is more difficult to set, but as a first guess
one can set it equal to the inverse ratio of the Einstein frequencies
corresponding to the intramolecular and quantum forces, that we calculate
during the course of the simulations. For particles confined in narrow tubes
we found that the optimal ratio $\Delta t / \delta t$ is of the order of 8-10.
There are at least two possible strategies for the parallelization of
the multiple time step algorithm for rigid rotors. In the first case, one can
distribute the number of beads among the available nodes, and in the second
case one can distribute the number of molecules.
Using the first method requires very frequent and very short messages to be
passed between the nodes, i.e. every short time step $\delta \tau$. In the
second case the quantum dynamics is performed locally, but the positions of
all the molecules have to be redistributed among the nodes at every long time
step in order to calculate the intermolecular forces: this results in less
frequent communications (i.e. every long time step $\Delta t$), but the amount
of data for a single communication is larger than in the first case.
In the case of non-interacting molecules (i.e. the zero pressure limit)
the last strategy results, of course, in a negligible amount of inter-node
communication and is therefore optimal.

\subsection{Calculation of the selectivity}

The selectivity of two components, say T$_2$ and H$_2$ is defined as:
\begin{equation}
S(T_2/H_2) = \frac {x_{T_2}/x_{H_2}}{y_{T_2}/y_{H_2}},
\label{eq:select}
\end{equation}
where $x$ and $y$ are the mole fractions in the adsorbed and bulk
phases, respectively.  

In the limit of zero pressure, when one can neglect the
adsorbate--adsorbate interaction, the selectivity depends only on the energy
levels $E^l$ of the adsorbed molecules, and can be written
as~\cite{challa,hathorn}
\begin{equation}
S_0(T_2/H_2) = \frac{Q_{H_2}^{\rm free}}{Q_{T_2}^{\rm free}}
\frac{Q_{T_2}}{Q_{H_2}}
= \left( \frac {m_{H_2}}{m_{T2}} \right)^{d/2}
\frac{Q_{T_2}^{\rm free-rot}}{Q_{H_2}^{\rm free-rot}}
\left[ \frac {\displaystyle\sum_{l}
\mathrm{exp}{\left( - E_{T_2}^l / k T \right)}}{\displaystyle\sum_{l}
\mathrm{exp}{\left( - E_{H_2}^l / k T \right)}} \right]
\label{eq:S0}
\end{equation}
where $Q^{\rm free}$ is the molecular partition function for the ideal gas,
$Q^{\rm free-rot}$ is the free rotor molecular partition function and $Q$ is
the molecular partition function for the given specie. We have denoted by $d$
the number of spatial dimensions in which confinement takes place. In the case
of hydrogen molecules in carbon nanotubes, $d=2$. In the zero pressure limit
the selectivity is a function of the energy levels of the two species, which
can be obtained by a direct diagonalization of the single-particle
Hamiltonian. We shall refer to this procedure as ``the simple model''.

We notice that, under the assumption that the rotational and translational
dynamics are independent, one can rewrite Eq.~(\ref{eq:S0}) as
\begin{equation}
  S_0(T_2/H_2) \simeq 
\left( 
\frac{m_{H_2}}{m_{T_2}} \frac{Q_{T_2}^{\rm tras}}{Q_{H_2}^{\rm tras}}
\right)
\left( 
\frac{Q_{T_2}^{\rm free-rot}}{Q_{H_2}^{\rm free-rot}}
\frac{Q_{T_2}^{\rm rot}}{Q_{H_2}^{\rm rot}}
\right)
= S_0^{\rm tras} S_0^{\rm rot}
\label{eq:S0_prod}
\end{equation}

The calculation of the selectivity of a confined T$_2$/H$_2$ can in principle
be performed using the definition given in Eq.~(\ref{eq:select}), evaluating
the mole fractions in the bulk and in the confined system using Grand
Canonical Monte Carlo simulations.

Since we expect quite high selectivities (possibly of the order of $10^5$ or
more), the evaluation of the mole fraction would require the simulation of
very large systems, of the order of at least $10^5$ molecules. In order to
avoid the use of such demanding calculations, we have employed the method
developed by Challa {\em et al.}~\cite{challa_zerop, pigcmc} to evaluate the
zero pressure selectivity.
With a straightforward extension to rigid rotors the selectivity in the limit
of zero pressure can be written as
\begin{equation}
  S_0({\mathrm T}_2/{\mathrm H}_2) = C ~ \langle \exp[-\beta \Delta U_{{\mathrm
  H}_2 \rightarrow {\mathrm T}_2}] \rangle_{{\mathrm H}_2}
  \label{eq:S0_pimc}
\end{equation}
where
\begin{equation}
  \Delta U_{{\mathrm H}_2 \rightarrow {\mathrm T}_2} = 
  \int_{m_{H2}}^{m_{T2}} dm ~ \left( \frac{d U_{int}}{d m} \right)
  \label{eq:integral}
\end{equation}
is the variation of the quantum potential energy when a H$_2$ molecule is
gradually transformed into a T$_2$ molecule by performing a number $N_{MC}$ of
Monte Carlo steps and the constant $C$ is given by
\begin{equation}
  C = \frac{Q^{\rm free-rot}(H_2)}{Q^{\rm free-rot}(T_2)}
  \left( \frac{m_{T_2}}{m_{H_2}} \right)^{\frac{3}{2} (P-1)}
  \left( \frac{I_{T_2}}{I_{H_2}} \right)^P
\end{equation}
where we have denoted by $I$ the inertia moment of a given specie, and
the average in Eq.~(\ref{eq:S0_pimc}) is performed on a simulation of the
lightest specie only.
A number of the order of $N_{MC} = 4000$ points are necessary to reach
convergence in the evaluation of the integral in Eq.~(\ref{eq:integral}) for
the H$_2$ to T$_2$ transformation.

In order to calculate the selectivity at finite pressures Challa {\em et
al.}~\cite{challa} have developed an efficient method to perform Grand
Canonical simulations of mixtures. The usual insertion and deletion moves are
performed for the heavier specie only (T$_2$ in our case), whereas the
lighter specie is inserted or deleted by performing T$_2
\leftrightarrow$~H$_2$ transformations.
A transformation move of a molecule of the specie 1 into a molecule of the
specie 2 is accepted with the probability
\begin{equation}
  {\cal P}_{1 \rightarrow 2} = {\rm min} \left[ 1,
    \frac{N_1}{N_2+1} 
    \left(\frac{\Lambda_1}{\Lambda_2}\right)^3 
    \exp[\beta (\mu_2 - \mu_1)] 
    \exp[-\beta \Delta U^{\rm ext}]
    \right] 
  \label{eq:alchemy}
\end{equation}
where $N$ is the number of molecules already present in the system, $\Lambda$
is the de~Broglie wavelength, $\mu$ the chemical potential (which is different
for the two species to account for a given bulk molar composition) and $\Delta
U^{\rm ext}$ is the difference of the sum of the fluid-fluid and solid-fluid
potential energies between the configurations $(N_1,N_2)$ and
$(N_1-1,N_2+1)$. Note that, in order to fulfill the detailed balance
condition, the probability of a T$_2 \rightarrow$~H$_2$ transformation attempt
must be equal to the probability of attempting the reverse move.

\subsection{Classical treatment of the rotational degrees of freedom}
\label{sec:class_rot}

In order to assess the importance of the quantized rotational degrees of
freedom in quantum sieving we now develop a formalism to describe a system in
which only the translational degrees of freedom are quantized and the
rotations are described classicaly.

In what follows we perform the derivation referring to a single rotor in an
external potential, in order to avoid a cumbersome notation. The extension to
interacting rotors is straightforward.

Consider a system whose Hamiltonian is given by $H = T(\hat{p}) + K(\hat{L}) +
V({\hat x},{\hat \Omega})$, where $T({\hat p}) = {\hat p}^2/2m$ is the kinetic
energy of translation, $K({\hat L}) = B {\hat L}^2$ is the kinetic energy of
rotation and $V({\hat x},{\hat \Omega})$ is the potential energy with a
dependence on the position operator $\hat x$ of the center of mass and the
direction $\hat \Omega$. The quantum mechanical partition function is given by
\begin{equation}
  Q = \int dx \sum_{l,m} \langle x;l,m| \exp[-\beta (T+K+V)] |x;l,m\rangle
\end{equation}

The classical treatment of some of the degrees of freedom correspond to the
assumption that the operators of the corresponding generalized coordinates and
momenta commute. Since we are interested in approximating the rotation as
classical, we proceed as if the rotational kinetic energy and the potential,
which depends on the molecular orientation, obey the commutation relation
$[{\hat V}, {\hat K}] = 0$, which implies $\exp[-\beta (T+V+K)] =
\exp[-\beta(T+V)] \exp[-\beta K]$. One can then perform the partial trace over
the rotational degrees of freedom, obtaining
\begin{eqnarray}
 q &=& {\rm Tr}_{rot} \exp[-\beta (T+K+V)] \nonumber \\
 &=& \sum_{lm} \langle l,m|\exp[-\beta (T+K+V)] | l,m \rangle \nonumber \\
 &=& \sum_{lm} \langle l,m|\exp[-\beta (T+V)] | l,m \rangle ~ e^{-
 \beta B l(l+1)} \nonumber \\
 &=& \int d\omega_1 d\omega_2 \sum_{lm} ~ \langle l m | \omega_1
 \rangle \langle \omega_1 | \exp[-\beta (T+V)] | \omega_2 \rangle
 \langle \omega_2 | l,m \rangle ~ e^{- \beta B l(l+1)} \nonumber\\
 &=& \int d\omega_1 d\omega_2 \sum_{lm} ~ Y^*_{lm}(\omega_1)
 Y_{lm}(\omega_2) \langle \omega_1 | \omega_2 \rangle \exp[-\beta
 (T+V(x,\omega_2)]~ e^{-\beta B l(l+1)}\nonumber \\
 &=& \int d\omega_1 
 \left (\sum_{l} \frac{2l+1}{4\pi} e^{-\beta B l(l+1)} \right)
 \exp[-\beta (T+V(x,\omega_1)] \nonumber \\
 &=& Q_{rot} \int \frac{d\omega}{4\pi} \exp[ -\beta
 (T(\hat{p})+V(\hat{x},\omega))]
\end{eqnarray}
where $Q_{rot}$ is the molecular partition function of the free rotor.
In the last expression the direction $\omega$ is the classical
direction of the rotor and $\hat{p}$ and $\hat{x}$ are the momentum
and position operators (still quantum mechanical).

When one applies the Trotter formula to the reduced density matrix
$q$ each of the beads corresponding to a given molecule has the
molecular axis pointing in the same direction as the others, since
$\omega$ is now considered as a classical variable.

\section{Results and discussion}
\label{sec:results}

\subsection{The potential model}

In order to assess the importance of quantized rotation on the selectivity, we
need a potential model that explicitely treats the hydrogen molecule as a
rigid rotor. To the best of our knowledge no such model has been extensively
tested in the literature. We have recently evaluated the zero pressure
selectivity of various potential models in the (3,6) nanotube in the framework
of the ``simple theory'', that is, using Eq.~(\ref{eq:S0}) with the energy
levels obtained by a direct diagonalization of the
Hamiltonian.~\cite{Garberoglio2006,goldfield2006} Many models predict very high
selectivities (in the range $10^7 - 10^{10}$), and we noticed a strong
dependence of the selectivity on the potential parameters.

We have chosen to use a reasonable potential model, well aware of the fact
that it might not be the best one to describe the actual interaction of
hydrogen molecules with themselves and with a carbon nanotube. We would
like to point out that our main interest in this work is to evaluate the
effect of quantized degrees of freedom on the selectivity, and not to develop
an accurate model to describe the hydrogen-carbon or hydrogen-hydrogen
interaction.

We describe the hydrogen molecule as a rigid rotor of length $l = 0.74$~\AA\
with two Lennard-Jones sites on the position of the hydrogen atoms, having as
parameters $\epsilon = 8.4$~K and $\sigma = 2.81$~\AA.~\cite{Murad78} The
Lennard-Jones radius of this model is less than the Lennard-Jones radius of
the Buch potential (where $\sigma_{\rm Buch} = 2.96$~\AA), which has been
shown to be a good spherical potential to describe the bulk properties of
hydrogen. The smaller $\sigma$ takes into account the fact that the potential
has two centers, separated by a distance corresponding to the actual gas phase
H$_2$ bond length.  The value of $\epsilon$ is almost one fourth of the
corresponding value for the Buch potential (where $\epsilon_{\rm Buch} =
34.2$~K) and this takes into account the fact that in a site-site model we
have actually four interactions between the two molecules.

We have also described the carbon atoms in the nanotubes with a Lennard-Jones
potential, using the Steele parameters $\sigma_C = 3.4$~\AA\ and $\epsilon_C =
28.0$~K.~\cite{Steele78} Solid-fluid interactions have been calculated using
the Lorentz--Berthelot mixing rules.
We have generated carbon nanotubes of various sizes and tabulated the
solid-fluid potential by averaging, in cylindrical coordinates, over the
length of a unit cell in the direction $z$ of the tube axis and over the angle
$\Phi$ in a plane orthogonal to the tube axis, 
thus obtaining the solid-fluid potential as a function of the distance of the
molecule's site from the nanotube axis only.

\begin{figure}
  \epsfig{file=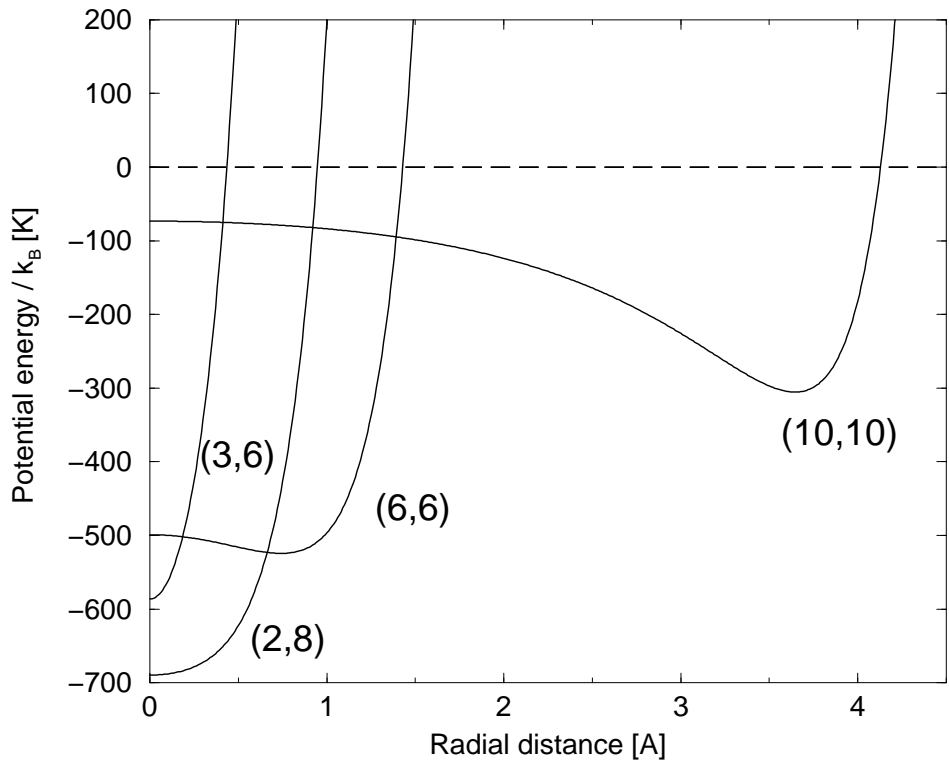,width=0.7\linewidth}
  \caption{The average potential energy curves for one of the hydrogen
  sites with the carbon nanotubes used in this study, as a function of the
  distance from the tube axis.}
  \label{fig:pot_shape}
\end{figure}

In this study we have focused our attention on the carbon nanotubes called
(3,6), (2,8), (6,6) and (10,10) with the standard nomencalture. These tubes
have geometrical radii of 3.1, 2.6, 4.1 and 5.1~\AA, respectively. We report
the profile of the potential energy with one of the hydrogen sites if
Fig.~\ref{fig:pot_shape}.

\subsection{The zero pressure selectivity}

\begin{table}
  \begin{tabular}{c|c|c|c|c}
    Simulation & Selectivity & Solid-fluid PE (K) & 
    Translational KE (K) & Rotational KE (K)\\
    \hline
    \hline
    (3,6)              & $6.34 \times 10^7 $ & -648 & 320 & 109 \\
    (3,6) classical    & 27000 & -900 & 280 & -- \\
    (3,6) simple model & 24400 & -932 & 271.5 & -- \\
    \hline
    (2,8)              & 52 & -1257 & 134 & 0 \\
    (2,8) classical    & 21 & -1297 & 115 & -- \\
    (2,8) simple model & 17 & -1313 &  98 & -- \\
    \hline
    (6,6)              & 2.3  & -1016 & 66 & 0   \\
    (6,6) classical    & 2.3  & -1018 & 55.7 & -- \\
    (6,6) simple model & 2.12 & -1022 & 41.9 & -- \\
    \hline
    (10,10)              & 5.6 & -524 & 71 & 0 \\
    (10,10) classical    & 4.2 & -532 & 66   & --\\
    (10,10) simple model & 4.7 & -556 & 58.1 & --\\
    \hline   
  \end{tabular}
  \caption{The values of the {\em para}-T$_2$/{\em para}-H$_2$ selectivities
  calculated for different nanotubes at 20~K. The potential and kinetic
  energies reported are those of the H${}_2$ molecule. The simple model
  results have been obtained by assuming that a molecule is aligned with the
  nanotube axis and that rotations do not contribute to the selectivity.}
  \label{tab:sel}
\end{table}

Our results for the influence of the quantization of the rotational degrees of
freedom on the selectivity are shown in Table~\ref{tab:sel}, where we report
zero pressure selectivities and average energies for hydrogen confined in the
(3,6), (2,8), (6,6) and (10,10) carbon nanotube.

The simulations have been performed using the method described in
Eq.~(\ref{eq:S0_pimc}). A system of at least 50 non interacting molecules was
equilibrated inside a carbon nanotube for at least 20000 HMC steps, and the
selectivity was then calculated by performing a production run of at least
20000 HMC moves. The selectivity was calculated from Eq.~(\ref{eq:S0_pimc}),
with configurations sampled every 50 HMC steps and using $N_{MC} = 4000$
points for the integration of Eq.~(\ref{eq:integral}). 

It is apparent that the explicit inclusion of the rotational degrees of
freedom has a dramatic effect on the selectivity, especially in the narrowest
tube, where it jumps from 27000 up to $6.34 \times 10^7$, a more than 2000 fold
increase. In the other tubes, the effect of the rotational degrees of
freedom is less dramatic, being of the order of 2.5 in the (2,8) tube and
almost negligible in the (6,6).

This enhancement is more than can be expected from a simple analysis that
assumes independence of the translational and rotational degrees of
freedom,~\cite{hathorn} which would give, in the case of the (3,6) tube with
our potential, a selectivity 750 times higher than the one obtained assuming
an isotropic potential.~\cite{Garberoglio2006} Rotational-translational
coupling
effects must be taken into account in narrow carbon nanotubes. We would also
like to point out that the actual magnitude of the rotational-translational
coupling on the selectivity calculated with the path integral simulations is
still higher than the one obtained using the approximate method that we have
developed in Ref.~\onlinecite{Garberoglio2006}, which would give, for the same
system 
analyzed here, an enhancement of the selectivity by a factor of 1650.

The physical origin of the higher selectivity due to the quantization of the
rotational degrees of freedom can be seen by analizying the difference between
the simulations in which the rotational are treated classically and the ones
in which the rotations are treated quantized.

In the presence of narrow confining potentials one expects to find the
molecules aligned with the nanotube axis, when rotations are treated
classically. In fact, we have calculated in Ref.~\onlinecite{Garberoglio2006}
the selectivity in the framework of the simple model assuming a perfect
alignment - so that the potential energy is a function of the center of mass
position only - and we can see, from the results reported in
Table~\ref{tab:sel}, that the results are in very good agreement with the path
integral simulations where rotational degrees of freedom are assumed to behave
classically.

\begin{figure}
  \epsfig{file=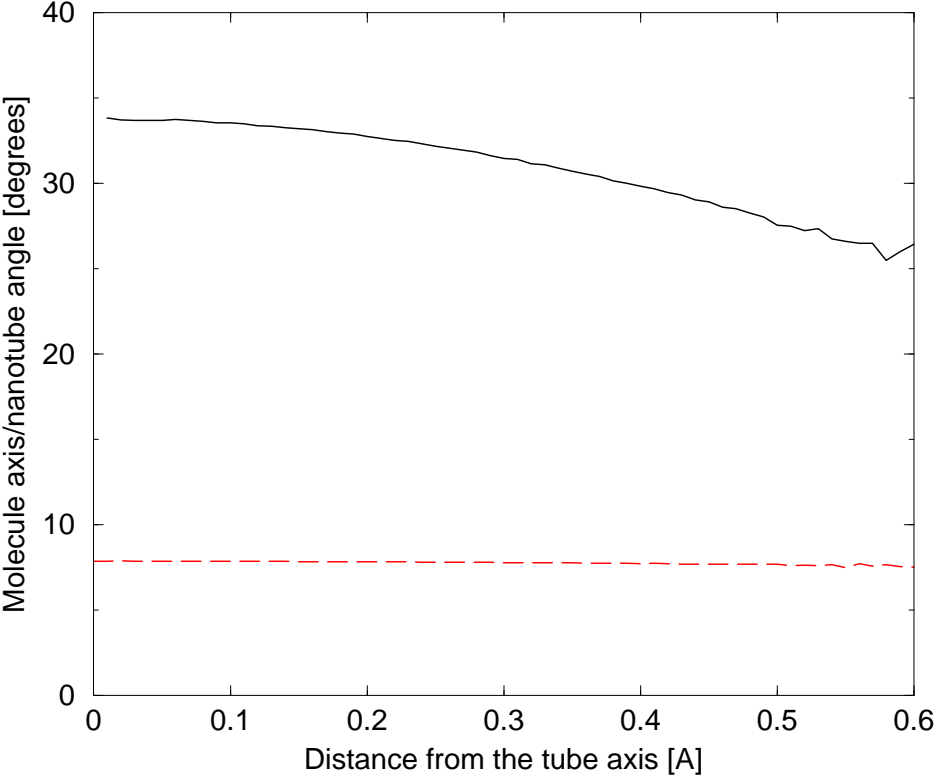,width=0.7\linewidth}
  \caption{The average angle between the nanotube and the molecular axes as a
  function of the distance from the (3,6) nanotube center for H$_2$. Solid
  line, full quantum simulations. Dashed line, classical treatment of the
  rotational degrees of freedom.}
  \label{fig:angle}
\end{figure}

We have also calculated the average angle $\Theta$ between the molecular axis
and the nanotube axis in the classical simulations as a function of the
distance of the molecule from the nanotube axis. One can see from
Fig.~\ref{fig:angle} that in the classical case one obtains $\Theta \simeq
8$~degrees, thus confirming an almost perfect alignement.

The situation changes dramatically upon quantization of the rotational degrees
of freedom, since the confining effect of the potential energy is now
counterbalanced by the quantum delocalization. The average angle $\Theta$
between the ``orientation'' of a given bead and the nanotube axis averages 
around 35 degrees at the center of the tube and has a slight decrease towards
25 degrees for the molecules which happen to be off center, where the
potential energy is steeper.

The effect of quantum fluctuations in the orientation can also be seen by the
high value of the rotational kinetic energy reported in
Table~\ref{tab:sel}. Hydrogen molecules confined in the narrowest tube have an
average rotational kinetic energy up to 109~K, a very high value when compared
to the free-rotor rotational energy at the same temperature, which is about
0.1~K due to the freezing of the rotational degrees of freedom on the $l=0$
spherically symmetric ground state.  Due to the confining potential that tends
to localize the molecular direction along the nanotube axis, the rotational
state of the adsorbed rotor is a superposition of higher angular momentum
states, resulting in a non-zero value of the average kinetic energy.

If we now consider, in a semiclassical picture, a molecule at a given distance
from the nanotube axis, we see that the quantization of the rotational degrees
of freedom (and the consequent rotational delocalization) has the effect that
the molecule samples region where the potential energy is higher with
respect to an almost perfectly aligned (classical) configuration. One can then
see that the average potential acting on the center of mass is steeper when
rotations are quantized than in the case when rotations are treated
classically, and that the steepness is higher for the lighter molecule than
for the heavier one.
As a consequence, the energy levels of the lighter rotors are more separated
with respect to the energy levels of the heavier specie, not only because of
a different mass, but also because the quantization of rotations has a
different effect on the two kind of molecules. In the light of the simple
theory formula, Eq.~(\ref{eq:S0}), we can then see that quantized rotations
enhance the spacing between the energy levels with a greater effect on the
light isotope, thus having a big effect on the selectivity, as is indeed
observed in the simulations.
This phaenomenon has been termed ``extreme two dimensional confinement'' by
Lu {\em et al.}~\cite{goldfield2006}
By a careful analysis of numerically exact eigenstates of hydrogen isotopes
confined in carbon nanotubes, they found that a large value of the selectivity
is to be expected in geometries so narrow that the rotational ground state
takes contribution from states with finite angular momentum. 
We have been able to show,~\cite{Garberoglio2006} using an approximate model
for the description of the coupled rotational and translational degrees of
freedom, that under these circumstances a very large contribution to the
selectivity does indeed come from the rotational degrees of freedom, as is
apparent in the exact result that we show in Table~\ref{tab:sel} for
the (3,6) tube.
In the other tubes the average rotational kinetic energy has been found to be
zero within the error bars, and the contribution of the rotational degrees of
freedom to the selectivity is correspondingly smaller.

\begin{figure}
  \epsfig{file=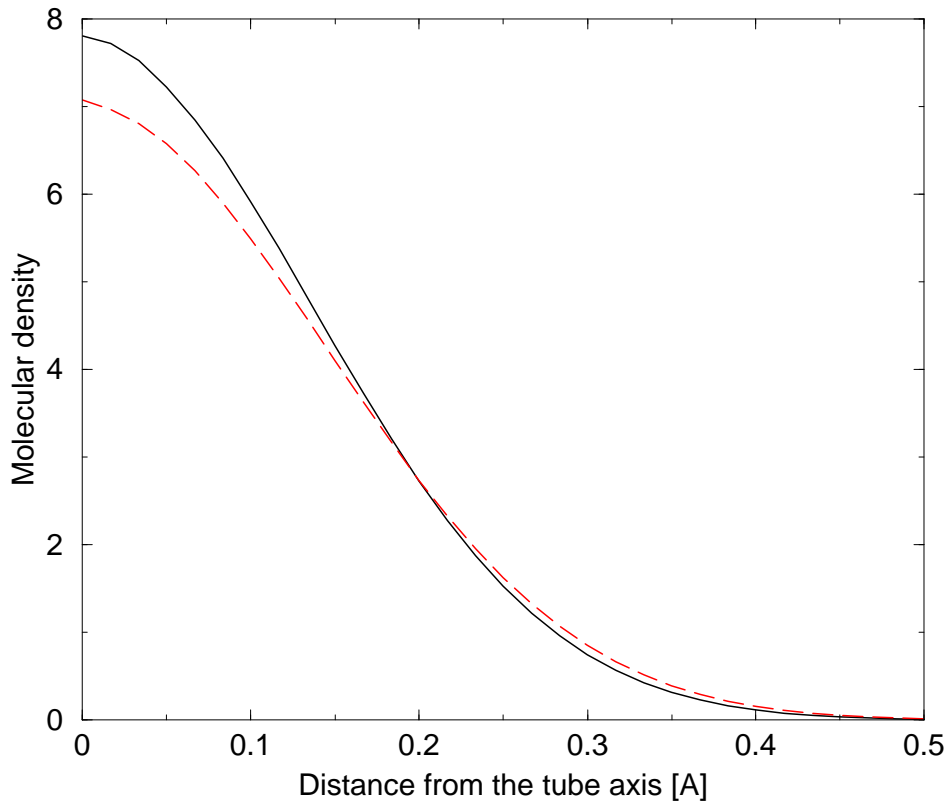,width=0.7\linewidth}
  \caption{Normalized densities for hydrogen confined in the (3,6) tube at
  20~K from Eq.~(\ref{eq:density}). Solid line, quantum treatment of the
  rotational degrees of freedom. Dashed line, classical treatment of the
  rotational degrees of freedom}
  \label{fig:density_3.6}
\end{figure}

One can also see the effect of the more confining effective potential by
plotting the expected density of a molecule as a function of the distance from
the tube center, when rotations are treated classically or quantized.
We show the results for hydrogen in the (3,6) tube in
Fig.~\ref{fig:density_3.6}, where it can be seen that the classical treatment
of the rotational degrees of freedom results in a lower density along the tube
axis, a signature of an effectively less confining external potential, as is
also apparent from the average kinetic and potential energies reported in
Table~\ref{tab:sel}: the quantized treatment of the rotational degrees of
freedom results in higher average kinetic and potential energies, when
compared to the case in which rotations are treated classically.

\subsection{Adsorption isotherms}

\begin{figure}
  \epsfig{file=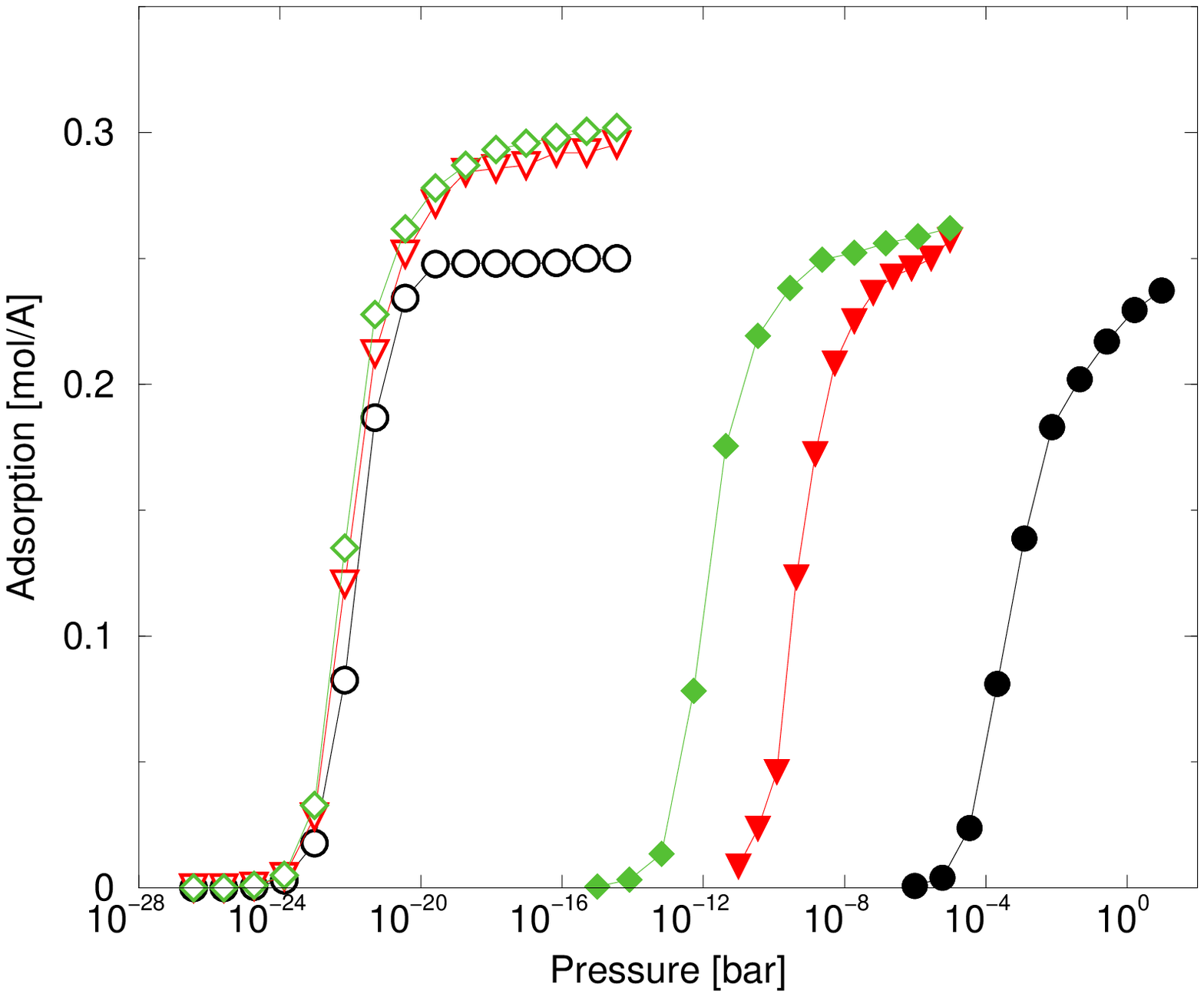,width=0.7\linewidth}
  \caption{Pure fluid adsorption isotherms for hydrogen isotopes in the (3,6)
  (filled symbols) and (6,6) (open symbols) carbon nanotubes at 20~K. The
  circles and diamonds refer to {\em para}-H${}_2$ and {\em para}-T${}_2$
  respectively, whereas the triangles refer to {\em ortho}-D${}_2$. The lines
  are drawn as a guide for the eye.}
  \label{fig:iso_36_66}  
\end{figure}

\begin{figure}
  \epsfig{file=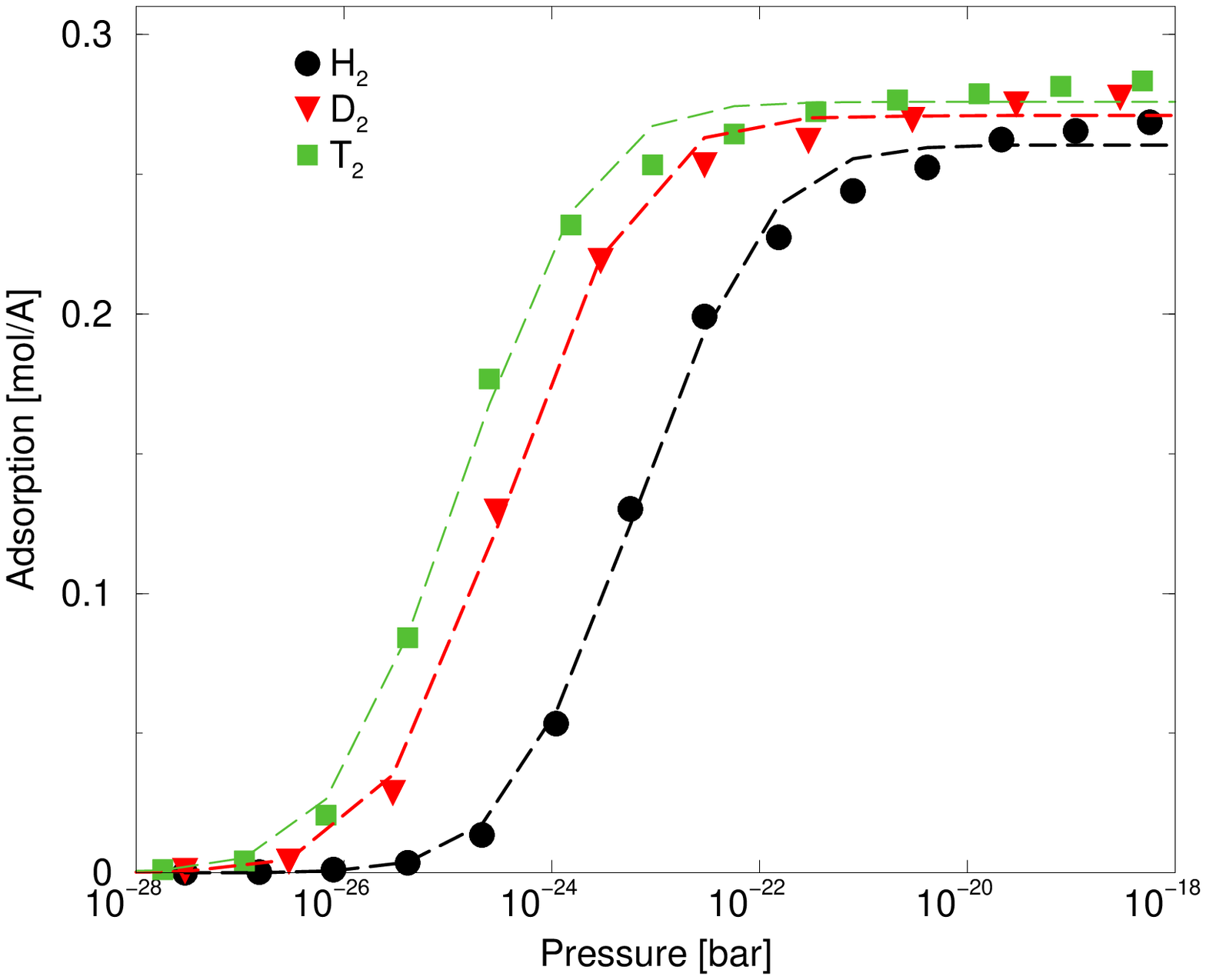,width=0.7\linewidth}
  \caption{Pure fluid adsorption isotherms for hydrogen isotopes in the (2.8)
  carbon nanotube at 20~K. The circles, triangles and diamonds refer to
  {\em para}-H${}_2$,  {\em ortho}-D${}_2$ and {\em para}-T${}_2$
  respectively. The lines are fit with the LUD isotherm.}
  \label{fig:iso_28}  
\end{figure}

Using the path integral simulations we have calculated the adsorption
isotherms for different hydrogen isotopes in the (3,6), (2,8) and (6,6) carbon
nanotubes. The results are shown in Figs.~\ref{fig:iso_36_66} and
\ref{fig:iso_28}.

As a general trend, we notice that the heavier specie is the one most readily
adsorbed, as is expected. Qualitatively, one can think that the larger thermal
de~Broglie wavelength of the lighter specie results in a larger effective
Lennard-Jones radius, thus hindering the adsorption.

Quantum effects result in a separation in pressure between the isotherms of
different isotopes. This is a negligible effect in large tubes, such as the
(6,6) where quantum mechanical effects do not influence the adsorption very
much (as apparent from the low selectivity). In this case the isotherms of the
different isotopes are very close, as can be seen in Fig.~\ref{fig:iso_36_66}.

The opposite is observed in the very confining (3,6) tube. It can be seen in
Fig.~\ref{fig:iso_36_66} that T${}_2$ can be adsorbed already at
$10^{-15}$~bar, whereas the adsorption of H${}_2$ is not detectable below
$10^{-5}$ bar, a difference of 10 orders of magnitude.

In the (2,8) tube we observe a sizeable presence of quantum effects, though
not as dramatic as in the (3,6), the isotherms of the lightest and heaviest
species are separated by two order of magnitude in pressure.

These isotherms allow us to clarify in detail how low the pressure must be
in order to fulfill the zero pressure limit discussed in the previous
section. For a given nanotube we can take as an estimate the pressure for
which the isotherm of the heavier specie starts to be significantly different
from zero.

We notice that a high zero pressure selectivity is correlated with the
separation in pressure of the isotherms corresponding to the two species. When
the heavier molecules begins to be adsorbed, the lighter one cannot easily do
so, hence the high values of the selectivity.

\subsection{Selectivity at finite pressures}

Using path integral simulations it is possible to investigate what happens to
the selectivity at finite pressures. We have used the very efficient method
of Eq.~(\ref{eq:alchemy}) to calculate the selectivity using the definition in
Eq.~(\ref{eq:select}).
Differently to what happens in the zero pressure limit, the selectivity at
finite pressure does depend on the assumed mole fraction in the bulk. We have
assumed that at the low pressures of our study, the chemical potentials
$\mu_1$ and $\mu_2$ of Eq.~(\ref{eq:alchemy}) can be evaluated using the ideal
gas expressions. They are therefore a function of the total pressure and the
assumed mole fraction in the bulk.

\begin{figure}
  \epsfig{file=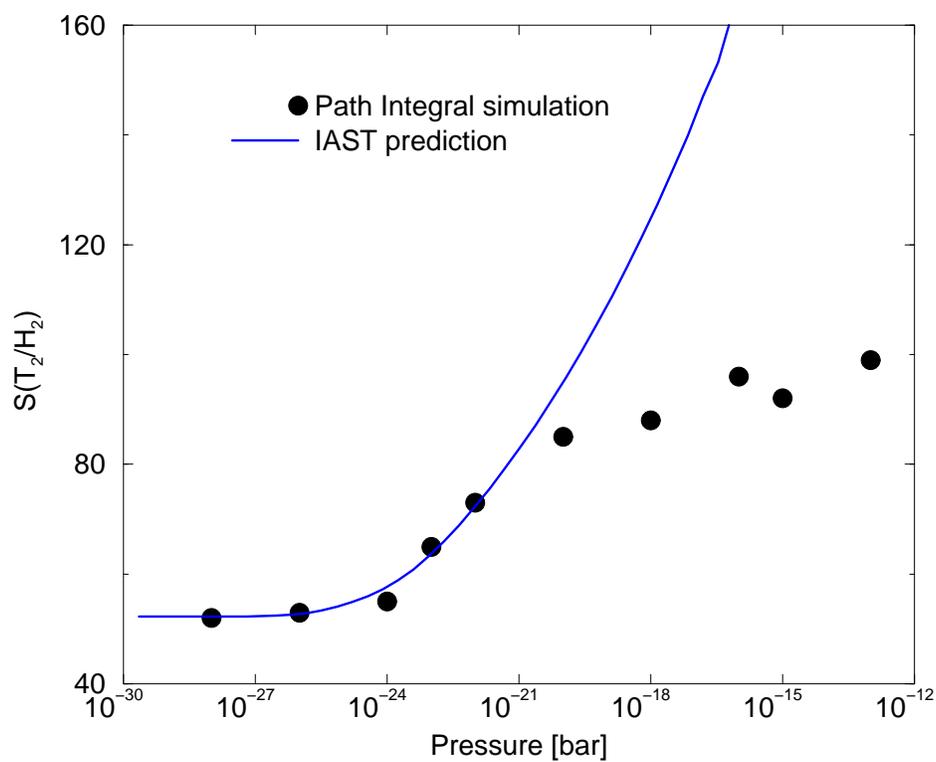,width=0.7\linewidth}
  \caption{Finite pressure T${}_2$/H${}_2$ selectivity in the (2,8)
  tube at 20~K, assuming a bulk mole fraction of T${}_2$ equal to $y_{T_2} =
  0.1$. Solid line, prediction from IAST theory. Circles, path integral
  simulations.}
  \label{fig:alchemy_28}  
\end{figure}

Since we are calculating the selectivity using the observed mole fraction in
the simulation box, it is apparent that we have to consider thermodynamic
conditions where we expect finite amount of both isotopes to be present in the
simulation box.
For a given pressure we can then fix the bulk mole fraction so that the two
species can be expected to be adsorbed more or less in the same amount. It is
clear that for a given zero pressure selectivity $S_0(A/B)$ bulk mole
fractions such that $y_A / y_B \simeq 1/S_0$ would result in an almost 1:1
ratio in the simulation cell. As a consequence less statistical errors can be
expected when evaluating the ensemble average of Eq.~(\ref{eq:select}). We
have decided to investigate the particular case of the (2,8) tube. Since
observed a value of $S_0({\rm T}_2/{\rm H}_2) = 52$ (see Table~\ref{tab:sel}),
we fixed the bulk mole fraction of T${}_2$ to the value $y_{T_2} = 0.1$. The
results are shown in Fig.~\ref{fig:alchemy_28}.

We have also evaluated the selectivity in the framework of the ideal adsorbed
solution theory (IAST), as outlined in Ref.~\onlinecite{challa}. Assuming
ideal solution behavior of the two species, it is possible to derive an
expression of the selectivity starting from the isotherms reported in
Fig.~\ref{fig:iso_28}. We have fitted our isotherms with the same functional
form used in Ref.~\onlinecite{challa}, and the quality of the fit can be seen
on the same figure.
The results of the IAST imply that the selectivity raises monotonically with
the pressure, and this prediction is related to the fact that upon
saturation the density of the heaviest specie is slightly larger than the
density of the heavier one.

Also the path integral simulations show an increase of the selectivity at
finite pressure, but this increase is not as steady as the IAST prediction. In
fact the selectivity seems to tend to a constant value as the pressure is
raised, as can be seen in Fig.~\ref{fig:alchemy_28}.
We notice that the correct selectivity starts to deviate from the approximate
IAST value at the pressure ($10^{-21}$~bar in this case) where the adsorption
isotherms begin to show saturation.

The reason of this behaviour can be traced back to the structure of the
adsorbed phase. Each of the two pure species occupy the same nanotube
with an almost equal linear density around $0.27$~molecules/\AA. Therefore one
might expect an ``average distance'' $L \simeq 3.7$~\AA\ between the molecules
in the case of a saturated mixture.

In going from zero to finite pressures one can then expect that the motion
along the $z$ coordinate of a given molecule is progressively hindered by the
ever close presence of the other ones, until a saturation condition is
reached, and further compression of the system becomes more difficult.

\begin{figure}
  \epsfig{file=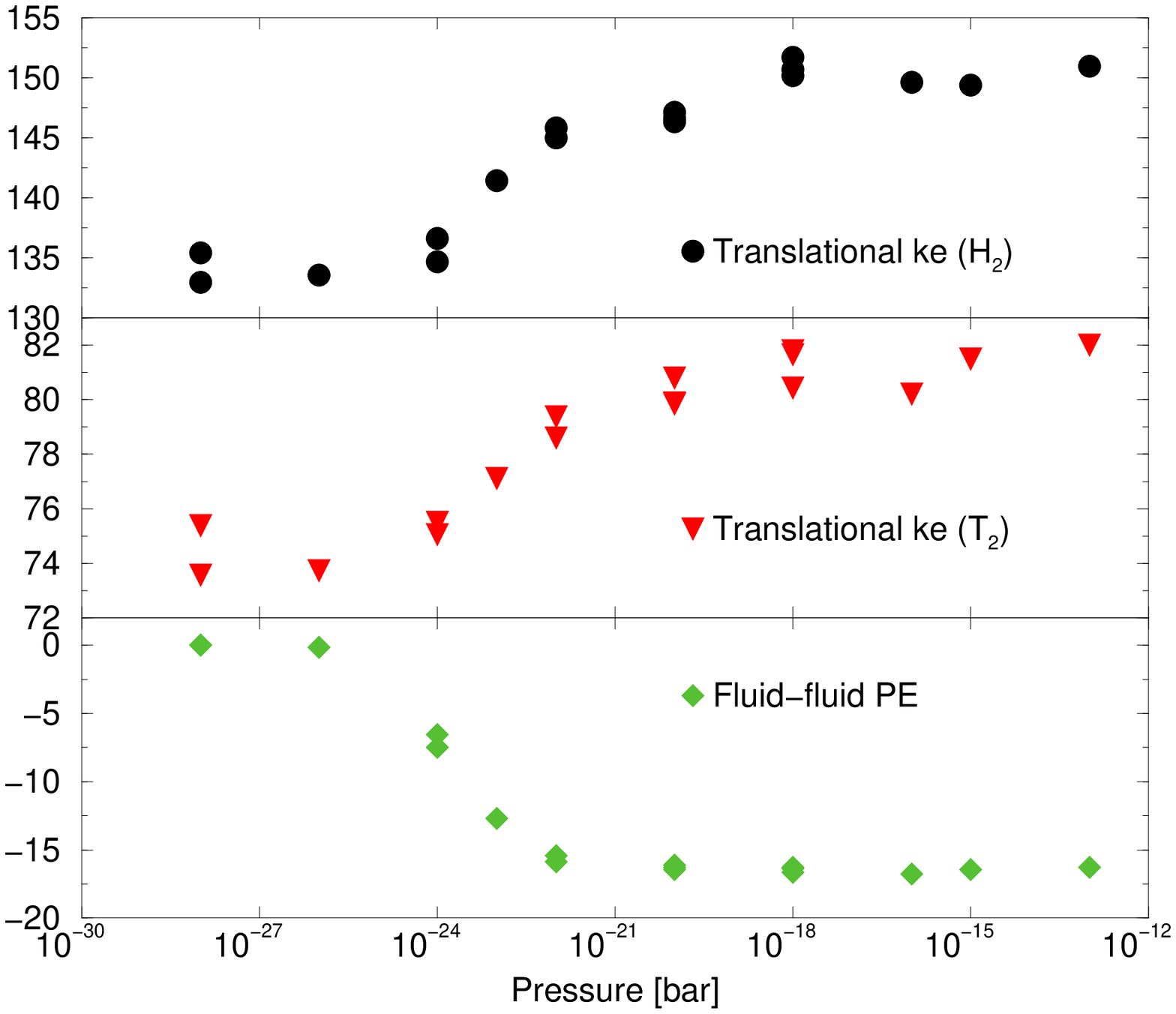,width=0.7\linewidth}
  \caption{Average translational kinetic energy (circles, H${}_2$; triangles,
  T${}_2$) and average solid fluid potential energy (diamonds) for the
  adsorption of a T${}_2$/H${}_2$ mixture ($y_{T_2} = 0.1$) in the (2,8)
  carbon nanotube at 20~K.}
  \label{fig:thermo_28}  
\end{figure}

This picture can be validated by the results reported in
Fig.~\ref{fig:thermo_28}, where we plot the average center of mass kinetic
energies for H${}_2$ and T${}_2$ as well as the average fluid-fluid energy for
the mixture adsorbed in the (2,8) tube.

We notice that the average potential energy per molecule tends to a constant
value of $V \simeq -16.5$~K, indicating that the molecular configuration does
not change very much after the onset of saturation.

The effect of the localization along the $z$ direction is apparent from
the behaviour of the kinetic energies as a function of the loading. The values
for T${}_2$ and H${}_2$, reported in Fig.~\ref{fig:thermo_28}, do indeed show
an increase at finite loading with respect to the zero pressure value. This
indicates a progressive hindrance of the motion along the nanotube axis, until 
the saturation point is reached and the kinetic energy tends to a constant
value.

It is possibile to make an estimate of these effects, by assuming that - at
saturation - all the molecules in the tube are separated from their neighbors
by the same distance, $L_{min}$. In order to calculate $L_{min}$ one can
assume that the molecules interact with an effective potential
obtained by averaging the original potential on the rotational degrees of
freedom. This takes into account the fact that the (2,8) tube is wide enough
so that the molecules are freely rotating.  
It is then possible to calculate the average potential energy per particle,
$V_{FF}(x)$, as a function of the distance $x$ between the molecules, assuming
that every molecule has two nearest neighbors located at $\pm x$.

The distance $L_{min}$ is then calculated as the point of mechanical
equilibrium, $\partial V_{FF}(x)/ \partial x = 0$, obtaining the value
$L_{min} = 3.43$~\AA, not too distant from the experimental average distance
given above. In this configuration the potential energy per particle is given
by $V_{FF}(L_{min}) = -51.6$~K.

We further assume that any molecule performs harmonic motions around the
potential energy minimum. In the actual system, of course, the dynamics is
determined by anharmonic effects, so that the following estimate has only a
heuristic value.  The ``spring constant'' for these oscillation is evaluated
to be $k_{HO} = \partial^2 V_{FF}/\partial x^2~(L_{min}) =
316.8$~K/\AA${}^2$. Hydrogen isotopes oscillating in this harmonic potential
would have a zero point energy of $E_0^{{\rm T}_2} = \hbar
\sqrt{k_{HO}/m_{{\rm T}_2}} /2 = 25.2$~K and $E_0^{{\rm H}_2} = 43.7$~K for
T${}_2$ and H${}_2$ respectively, so that we might expect $V > V_{FF}(L_{min})
+ E_0^{{\rm T}_2} = -26.4$~K, close to the actual value observed.

One can further estimate the asymptotic value of the selectivity at finite
pressure by assuming the that its value at saturation is given by the product
of the zero pressure value $S_0$ and the selectivity $S_{HO}$ due to the
effective 1D harmonic oscillator discussed above. Using and independent
particle approach,~\cite{wang,challa_zerop} the value of $S_{HO}$
corresponding to T${}_2$ and H${}_2$ in a harmonic oscillator of spring
constant $k_{HO}$ turns out to be $S_{HO} = 1.6$, in reasonable agreement with
the observed behavior of the selectivity which tends, at saturation, to twice
the zero pressure value.

\section{Conclusions}

In this paper we have presented an algorithm for the hybrid path integral
Monte Carlo simulation of rigid diatomic molecules. 
We show how to calculate torques from the expression of the rotational density
matrix and we moreover show an approximate expression for the rotational
density matrix of heteronuclear molecules that, in the large $P$ limit, is
completely analogous to the expression for the translational degrees of
freedom.
We have developed a velocity Verlet like integrator for the rotational degrees
of freedom suitable for a multiple time step approach.

We have then applied this method to the calculation of the {\em
para}-T${}_2$/{\em para}-H${}_2$ selectivity in various carbon nanotubes at
20~K. We have discussed, in particular, the effect of the quantized rotational
degrees of freedom on the selectivity by developing a simulation method that
allows a classical treatement of the rotations while keeping a quantum
tratment of the translational degrees of freedom.

We show that the explicit inclusion of quantized rotations enhances the
zero pressure selectivity by a factor of more than $2000$ in the narrowest
tube we have investigated (the (3,6)), but does not have such a dramatic
effect in larger tubes. The rotational degrees of freedom contribute by less
than a factor of two in the (2,8) and larger tubes.

We have been able to investigate the effect of finite pressures on the
adsorption and the selectivity. We have calculated the adsorption isotherms of
various hydrogen isotopes in different tubes at 20~K and shown that quantum
effects hinder the adsorption of the lighter specie, whose isoterm can be
separated by many order of magnitude in pressure from the one of the heavier
specie.

We have been able to calculate pressure dependence on the selectivity on the
pressure and found that, for a 0.1 molar mixture of T${}_2$ in the (2,8) tube,
the selectivity tends to twice its zero pressure value when the pressure is
raised. We correlate this behaviour to the hindrance of the molecular motion
along the nanotube axis.

\newpage 

\appendix

\section{Velocity-Verlet integrator for rigid linear molecules}
\label{sec:appendix}

The dynamics of a rigid rotor is described by the equations
\begin{eqnarray}
  \frac{d {\mathbf e}}{d t} &=& {\mathbf \varpi} \times {\bf e} \label{eq:e}\\
  I \frac{d {\mathbf \varpi}}{d t} &=& N \label{eq:w}
\end{eqnarray}
where ${\bf e}$ is a unit vector in the direction of the rotor and,
$\varpi$ is the angular velocity, $I$ is the inertia moment and $N$ is
the torque applied to the system.

These dynamical variable are redundant, since it can be seen that the
norm of the vector ${\bf e}$ is a constant of the motion described by
the previous equations. 
Since the torque $N$ is, by construction, always orthogonal to the
axis vector, the component of the angular velocity along the unit
vector is also a constant of motion, and is usually set as zero.

The previous equations cannot be then put in the form of a Hamiltonian
system. In order to develop a time-reversible integrator that can be
used in the hybrid Monte Carlo method, we demonstrate that it is
indeed possibile to integrate the equations using a velocity Verlet
like, adapted to take into account the abovementioned constraint.
The multi-step algorithm can them be developed by analogy to the
velocity Verlet case.

Using the Taylor expansion we can write
\begin{eqnarray}
  {\bf e}(t + \delta t) &\simeq& {\bf e}(t) + \delta t  \frac{d {\bf e}}{d t}
  + \frac{1}{2} (\delta t)^2 \frac{d^2 {\bf e}}{dt^2}  \\
 &=& {\bf e} + \delta t (\varpi \times {\bf e}) + \frac{1}{2}
  (\delta t)^2 \frac{d}{dt} (\varpi \times {\bf e}) \\
 &=& {\bf e} + \delta t (\varpi \times {\bf e}(t)) + \frac{1}{2}
 (\delta t)^2 \left(
  \frac{d \varpi}{dt} \times {\bf e} + \varpi \times (\varpi
  \times {\bf e}) \right) \\
 &=& {\bf e}(t) + \delta t \left[ \left(\varpi(t) + \frac{\delta t}{2}
  \frac{N(t)}{I} \right) \times {\bf e}(t) \right] - \frac{(\delta
  t)^2}{2} \varpi^2 {\bf e}(t) \label{eq:eevol}
\end{eqnarray}
where we have restored the explicit time dependence on time in the
last passage.

The last term in Eq.~(\ref{eq:eevol}) assures that the length of the
unit vector describing the direction of the rotor remains fixed. In
our code we re-normalize the unit vector after each time step.

The equation for the angular velocity becomes
\begin{eqnarray}
  \varpi(t + \delta t) &\simeq& \varpi(t) + \delta t \frac{d \varpi}{d t} +
  \frac{(\delta t)^2}{2} \frac{d^2 \varpi}{d t^2} \\
  &=& \varpi + \frac{\delta t}{2} \dot{\varpi} + \frac{\delta t}{2}
  \left( \dot\varpi + \delta t \ddot \varpi \right) \\
  &=& \varpi(t) + \frac{\delta t}{2} \frac{N(t)}{I} + \frac{\delta
  t}{2} \frac{N(t+\delta t)}{I} \label{eq:wevol}
\end{eqnarray}

So that one can construct a velocity Verlet like algorithm for the
rotational degrees of freedom as
\begin{enumerate}
  \item{Calculate the angular velocity at half time step $\varpi(t + \delta
  t/2) =\varpi(t) + \frac{\delta t}{2} \frac{N(t)}{I}$}
  \item{Advance the orientation ${\mathbf e}$ at full time step, using
  Eq.~(\ref{eq:eevol})}
  \item{Calculate the torques at the time $t+\delta t$}
  \item{Advance the angular velocity at full time step}
\end{enumerate}

\section{Approximate treatment of the rotational density matrix}
\label{sec:app_rot}

It is known to be impossible to rewrite the rotational partition
function in Eq.~(\ref{eq:rot_dm}) as the partition function of an effective
harmonic potential, as happens for the translational degrees of freedom.

We have noticed, though, that if $P$ is large enough the rotational partition
function of a heteronuclear molecule can be approximated by an expression
analogous to Eq.~(\ref{eq:tke}) for the translational degrees of freedom.

\begin{equation}
  \Xi_{ij}({\theta}) = 
  \sum_{J=0}^{\infty} \frac{2J+1}{4 \pi} P_J(\cos\theta_{ij})
  \exp[-\beta J(J+1) B/P] \mathop\rightarrow_{P \rightarrow \infty}
  A \exp\left[ -\frac{\beta K}{2} \theta^2_{ij} \right]
\label{eq:rot_dm_approx}
\end{equation}
where
\begin{eqnarray}
  A &=& \frac{I k_B T P}{2 \pi \hbar^2} \label{eq:A}\\
  K &=& \frac{I P (k_B T)^2}{\hbar^2} \label{eq:K}
\end{eqnarray}

Eq.~(\ref{eq:A}) can be demonstrated by setting $\theta_{ij} = 0$ and
approximating the sum over $J$ in Eq.~(\ref{eq:rot_dm}) with an integral. We
have not been able to demonstrate Eq.~(\ref{eq:rot_dm_approx}) analytically,
but we have verified that the first 20 coefficients of the expansion in
$\theta_{ij}$ of both sides of Eq.~(\ref{eq:rot_dm_approx}) are the same in
the $P \rightarrow \infty$ limit and that the exact and approximate density
matrices are very similar in the same limit.
We would like to point out the formal similarity between
Eqs~(\ref{eq:A}) and (\ref{eq:K}) and the corresponding ones for the
translational degrees of freedom, Eqs.~(\ref{eq:a}) and (\ref{eq:kappa}): in
the large $P$ limit the quantum mechanical effect corresponds to the action of
an harmonic torque between the rotors in adjacent time slices.

\begin{figure}
  \epsfig{file=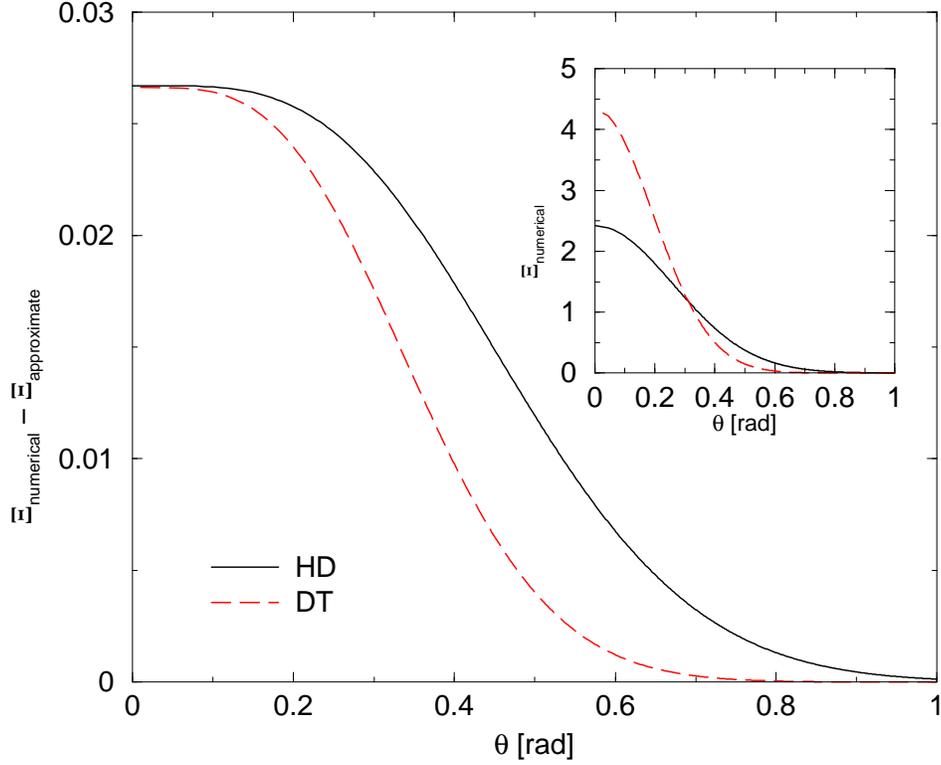,width=0.7\linewidth}
  \caption{The difference between the approximated and the numerical
  rotational density matrix for HD (solid line) and DT (dashed line) at
  20~K with $P=100$. The numerical density matrices for the same Trotter
  numbers are shown in the inset.}
  \label{fig:rotdm}
\end{figure}

We show in Fig.~\ref{fig:rotdm} the difference between the numerical and the
approximated rotational density matrix of Eq.~(\ref{eq:rot_dm_approx}), for HD
and DT at a temperature $T=20$~K, using a Trotter number $P=100$. We can see
that the approximate expression is accurate within 1.5\% in 
the low $\theta$ region for the ligher specie. At a constant Trotter number,
the approximation is of course better for a heavier molecule such as DT.

One can see from Eq.~(\ref{eq:rot_dm_approx}) that in the large $P$ limit the
quantum rotational density matrix is analogous to the translational
case (see Eq.~(\ref{eq:tke})), i.e. represents an effective harmonic potential
energy between adjacent beads in the classical mapping.

\begin{figure}
  \epsfig{file=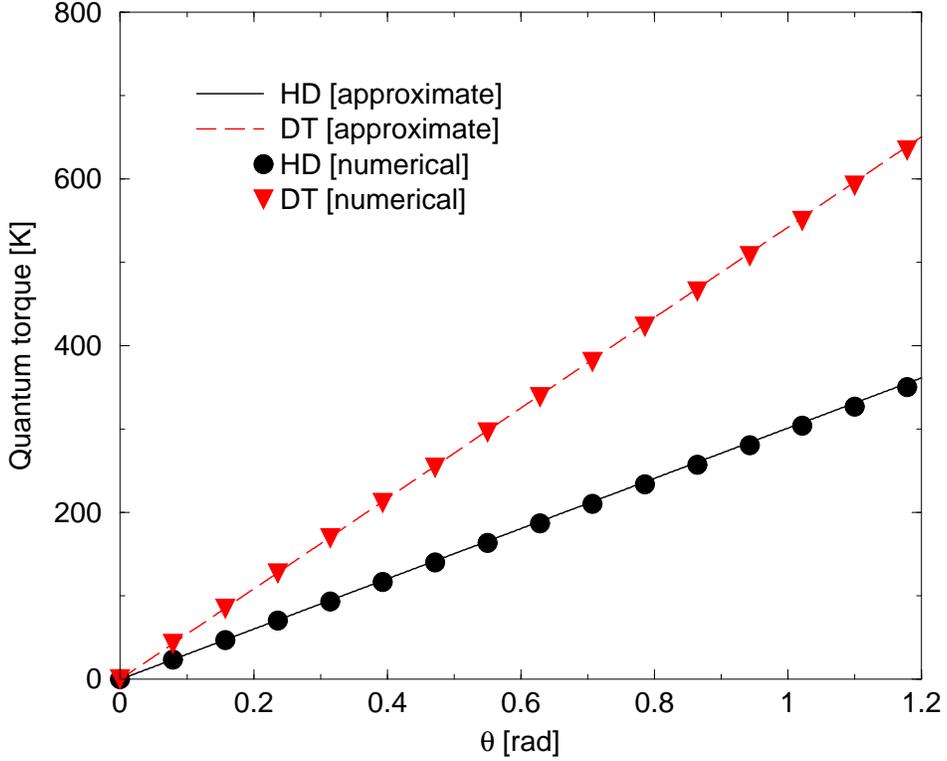,width=0.7\linewidth}
  \caption{The quantum torque calculated numerically using
  Eq.~(\ref{eq:qtorque}) (symbols) and analytically using the ``spring
  constant'' given in Eq.~(\ref{eq:K}) (lines). Solid line and circles refer to
  HD at 20K, dashed line and triangles to DT at 20K. In both cases we have set
  $P=100$.}
  \label{fig:app_torque}
\end{figure}

This is also apparent in the calculation of the torques, that we use in the
hybrid Monte Carlo calculation. We report in Fig.~\ref{fig:app_torque} the
results obtained using the numerical method derived in Eq.~(\ref{eq:qtorque})
and the analytical calculation, using the ``spring constant'' of
Eq.~(\ref{eq:K}). The two methods give virtually the same values, even though
the numerical calculation shows numerical instabilities for angles just a
little higher than the one shown in Fig.~\ref{fig:app_torque}.

Unfortunately the approximation of Eq.~(\ref{eq:rot_dm_approx}) is not
applicable for ortho- and para- species, because $\Xi_{ij}(\theta)$ does
not fall monotonously to zero for large $\theta$, as implied by the
right hand side of Eq.~(\ref{eq:rot_dm_approx}). 

\begin{figure}
  \epsfig{file=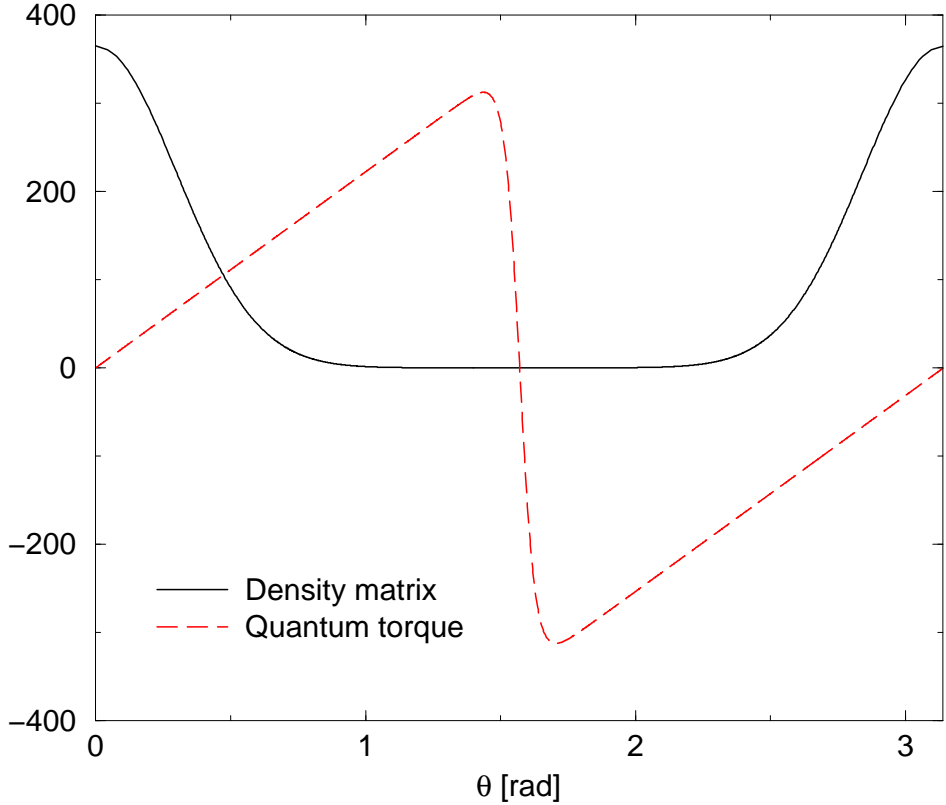,width=0.7\linewidth}
  \caption{Density matrix (solid line) and corresponding torque (dashed line)
  for {\em para}-H${}_2$ at 20K, using $P=100$.}
  \label{fig:dm}
\end{figure}

If the summation in Eq.~(\ref{eq:rot_dm_approx}) is restricted to the even
angular momenta only, one obtains a density matrix with the propriety
$\Xi_{ij}(\pi/2 + \theta) = \Xi_{ij}(\pi/2 - \theta)$, such as the
one shown in Fig.~\ref{fig:dm}. ($\Xi_{ij}(\theta)$ would be odd with
respect to the point $\theta = \pi/2$ if the odd angular momenta only were
used in Eq.~(\ref{eq:rot_dm_approx}).)

The approximation of Eq.~(\ref{eq:rot_dm_approx}) is still valid for $\theta <
\pi/2$, with the modification $A \rightarrow A/2$. Since the density matrix is
even with respect to $\theta = \pi/2$, one could try to approximate it as

$$
\Xi(\theta_{ij}) \simeq \frac{A}{2} \exp\left[-\frac{K \theta_{ij}^2}{2 k_B
    T}\right] \Theta(\pi/2 - x) + \frac{A}{2} \exp\left[-\frac{K
    (\pi - \theta_{ij})^2}{2 k_B T}\right] \Theta(x - \pi/2)
$$
where we have denoted by $\Theta(x)$ the step function, which is 0 for $x<0$
and 1 for $x>0$. Unfortunately this will result in a discontinuous quantum
torque at $\theta = \pi/2$, and we have not pursued the idea
further. Nonetheless the approximate expression of
Eqs.~(\ref{eq:rot_dm_approx}) -- (\ref{eq:K}) may prove to be useful for an
efficient quantum simulation of heteronuclear molecules, where no such
discontinuities appear.

It is interesting to notice that in the classical limit $\hbar \rightarrow 0$
the torque constant of Eq.~(\ref{eq:K}) goes to infinity. 
One could then conclude, albeit heuristically, that a classical treatment of
the rotational degrees of freedom only can be done by assuming that the
orientations of all the beads corresponding to one molecule all point in
the same direction, in agreement with the derivation presented in
Sec.~\ref{sec:class_rot}.

In order to test the validity of the approximation in
Eq.~(\ref{eq:rot_dm_approx}), we have evaluated the zero pressure selectivity
of a DT/HT mixture in the (3,6) tube using the approximate and exact method to
evaluate the rotational density matrix.
The results, shown in Table~\ref{tab:approx} show that the approximate method
compares very well with the exact one, even if the rotational kinetic energies
are sensibly different (but still within the quite large error bars) and the
zero pressure selectivity of a DT/HD mixture is overestimated by 50\%. In this
type of calculation, which considered $N=64$ particles with $P=100$ and
parallelization on 8 nodes, we have observed a 35\% speedup using the
approximate method.

\begin{table}
  \begin{tabular}{c|cc}
    Property & Exact result & Approximate result \\
    \hline
    Solid-fluid potential energy (K)& $-745 \pm 5$ & $-757 \pm 5$ \\
    Translational kinetic energy (K)& $252.8 \pm 0.5$ & $253.2 \pm 0.5$ \\
    Rotational kinetic energy (K)   & $172 \pm 43$ & $192 \pm 44$ \\
    $\langle \Delta \theta \rangle$ (degrees) & $17.5 \pm 0.2$ & $17.4 \pm 0.2$ \\
    $R_{gyr}$ (\AA)                & $0.314 \pm 0.008$ & $0.316 \pm 0.008$\\
    $S_0({\rm DT}/{\rm HD})$ & 5450 & 8470 \\

  \end{tabular}
\caption{Comparison of the approximate and exact method for the calculation of
  the density matrix. Properties of HD in a (3,6) carbon nanotube at 20~K with
  $P=100$. $\langle \Delta \theta \rangle$ is the average angle between the
  orientation of two adjacent beads and $R_{gyr}$ is the gyration radius of
  the classical ring polymer.}
\label{tab:approx}
\end{table}

\bibliography{sieving}

\begin{thebibliography}{21}
\expandafter\ifx\csname natexlab\endcsname\relax\def\natexlab#1{#1}\fi
\expandafter\ifx\csname bibnamefont\endcsname\relax
  \def\bibnamefont#1{#1}\fi
\expandafter\ifx\csname bibfnamefont\endcsname\relax
  \def\bibfnamefont#1{#1}\fi
\expandafter\ifx\csname citenamefont\endcsname\relax
  \def\citenamefont#1{#1}\fi
\expandafter\ifx\csname url\endcsname\relax
  \def\url#1{\texttt{#1}}\fi
\expandafter\ifx\csname urlprefix\endcsname\relax\def\urlprefix{URL }\fi
\providecommand{\bibinfo}[2]{#2}
\providecommand{\eprint}[2][]{\url{#2}}

\bibitem[{\citenamefont{Katorski and White}(1964)}]{KatorskiW64}
\bibinfo{author}{\bibfnamefont{A.}~\bibnamefont{Katorski}} \bibnamefont{and}
  \bibinfo{author}{\bibfnamefont{D.}~\bibnamefont{White}}, \bibinfo{journal}{J.
  Chem. Phys.} \textbf{\bibinfo{volume}{40}}, \bibinfo{pages}{3183}
  (\bibinfo{year}{1964}).

\bibitem[{\citenamefont{Moiseyev}(1975)}]{Moiseyev75}
\bibinfo{author}{\bibfnamefont{N.}~\bibnamefont{Moiseyev}},
  \bibinfo{journal}{J. Chem. Soc. Faraday Trans. 1}
  \textbf{\bibinfo{volume}{71}}, \bibinfo{pages}{1830} (\bibinfo{year}{1975}).

\bibitem[{\citenamefont{Beenakker et~al.}(1995)\citenamefont{Beenakker, Borman,
  and Krylov}}]{krylov}
\bibinfo{author}{\bibfnamefont{J.}~\bibnamefont{Beenakker}},
  \bibinfo{author}{\bibfnamefont{V.}~\bibnamefont{Borman}}, \bibnamefont{and}
  \bibinfo{author}{\bibfnamefont{S.}~\bibnamefont{Krylov}},
  \bibinfo{journal}{Chem. Phys. Lett.} \textbf{\bibinfo{volume}{232}},
  \bibinfo{pages}{379} (\bibinfo{year}{1995}).

\bibitem[{\citenamefont{Wang et~al.}(1999)\citenamefont{Wang, Challa, Sholl,
  and Johnson}}]{wang}
\bibinfo{author}{\bibfnamefont{Q.}~\bibnamefont{Wang}},
  \bibinfo{author}{\bibfnamefont{S.}~\bibnamefont{Challa}},
  \bibinfo{author}{\bibfnamefont{D.}~\bibnamefont{Sholl}}, \bibnamefont{and}
  \bibinfo{author}{\bibfnamefont{J.}~\bibnamefont{Johnson}},
  \bibinfo{journal}{Phys. Rev. Lett.} \textbf{\bibinfo{volume}{82}},
  \bibinfo{pages}{956} (\bibinfo{year}{1999}).

\bibitem[{\citenamefont{Challa et~al.}(2001)\citenamefont{Challa, Sholl, and
  Johnson}}]{challa_zerop}
\bibinfo{author}{\bibfnamefont{S.}~\bibnamefont{Challa}},
  \bibinfo{author}{\bibfnamefont{D.}~\bibnamefont{Sholl}}, \bibnamefont{and}
  \bibinfo{author}{\bibfnamefont{J.}~\bibnamefont{Johnson}},
  \bibinfo{journal}{Phys. Rev. B} \textbf{\bibinfo{volume}{63}},
  \bibinfo{pages}{245419} (\bibinfo{year}{2001}).

\bibitem[{\citenamefont{Challa et~al.}(2002)\citenamefont{Challa, Sholl, and
  Johnson}}]{challa}
\bibinfo{author}{\bibfnamefont{S.}~\bibnamefont{Challa}},
  \bibinfo{author}{\bibfnamefont{D.}~\bibnamefont{Sholl}}, \bibnamefont{and}
  \bibinfo{author}{\bibfnamefont{J.}~\bibnamefont{Johnson}},
  \bibinfo{journal}{J. Chem. Phys.} \textbf{\bibinfo{volume}{116}},
  \bibinfo{pages}{814} (\bibinfo{year}{2002}).

\bibitem[{\citenamefont{Hathorn et~al.}(2001)\citenamefont{Hathorn, Sumpter,
  and Noid}}]{hathorn}
\bibinfo{author}{\bibfnamefont{B.}~\bibnamefont{Hathorn}},
  \bibinfo{author}{\bibfnamefont{B.}~\bibnamefont{Sumpter}}, \bibnamefont{and}
  \bibinfo{author}{\bibfnamefont{D.}~\bibnamefont{Noid}},
  \bibinfo{journal}{Phys. Rev. A} \textbf{\bibinfo{volume}{64}},
  \bibinfo{pages}{022903} (\bibinfo{year}{2001}).

\bibitem[{\citenamefont{Trasca et~al.}(2003)\citenamefont{Trasca, Kostov, and
  Cole}}]{trasca}
\bibinfo{author}{\bibfnamefont{R.}~\bibnamefont{Trasca}},
  \bibinfo{author}{\bibfnamefont{M.}~\bibnamefont{Kostov}}, \bibnamefont{and}
  \bibinfo{author}{\bibfnamefont{M.}~\bibnamefont{Cole}},
  \bibinfo{journal}{Phys. Rev. B} \textbf{\bibinfo{volume}{67}},
  \bibinfo{pages}{035410} (\bibinfo{year}{2003}).

\bibitem[{\citenamefont{Lu et~al.}(2003)\citenamefont{Lu, Goldfield, and
  Gray}}]{goldfield}
\bibinfo{author}{\bibfnamefont{T.}~\bibnamefont{Lu}},
  \bibinfo{author}{\bibfnamefont{E.}~\bibnamefont{Goldfield}},
  \bibnamefont{and} \bibinfo{author}{\bibfnamefont{S.}~\bibnamefont{Gray}},
  \bibinfo{journal}{J. Phys. Chem. B} \textbf{\bibinfo{volume}{107}},
  \bibinfo{pages}{12989} (\bibinfo{year}{2003}).

\bibitem[{\citenamefont{Garberoglio et~al.}(2006)\citenamefont{Garberoglio,
  DeKlavon, and Johnson}}]{Garberoglio2006}
\bibinfo{author}{\bibfnamefont{G.}~\bibnamefont{Garberoglio}},
  \bibinfo{author}{\bibfnamefont{M.}~\bibnamefont{DeKlavon}}, \bibnamefont{and}
  \bibinfo{author}{\bibfnamefont{J.}~\bibnamefont{Johnson}},
  \bibinfo{journal}{J. Phys. Chem. B} \textbf{\bibinfo{volume}{110}},
  \bibinfo{pages}{1733} (\bibinfo{year}{2006}).

\bibitem[{\citenamefont{Lu et~al.}(2006)\citenamefont{Lu, Goldfield, and
  Gray}}]{goldfield2006}
\bibinfo{author}{\bibfnamefont{T.}~\bibnamefont{Lu}},
  \bibinfo{author}{\bibfnamefont{E.}~\bibnamefont{Goldfield}},
  \bibnamefont{and} \bibinfo{author}{\bibfnamefont{S.}~\bibnamefont{Gray}},
  \bibinfo{journal}{J. Phys. Chem. B} \textbf{\bibinfo{volume}{110}},
  \bibinfo{pages}{1742} (\bibinfo{year}{2006}).

\bibitem[{\citenamefont{Landau and Binder}(2000)}]{binder}
\bibinfo{author}{\bibfnamefont{D.}~\bibnamefont{Landau}} \bibnamefont{and}
  \bibinfo{author}{\bibfnamefont{K.}~\bibnamefont{Binder}},
  \emph{\bibinfo{title}{A guide to Monte Carlo simulations in statistical
  physics}} (\bibinfo{publisher}{Cambridge University Press},
  \bibinfo{year}{2000}), chap.~\bibinfo{chapter}{8}.

\bibitem[{\citenamefont{Marx and M{\"u}ser}(1999)}]{rotors}
\bibinfo{author}{\bibfnamefont{D.}~\bibnamefont{Marx}} \bibnamefont{and}
  \bibinfo{author}{\bibfnamefont{M.}~\bibnamefont{M{\"u}ser}},
  \bibinfo{journal}{J. Phys.: Condens. Matter} \textbf{\bibinfo{volume}{11}},
  \bibinfo{pages}{R117} (\bibinfo{year}{1999}).

\bibitem[{\citenamefont{Cui et~al.}(1997)\citenamefont{Cui, Cheng, Adler, and
  Whaley}}]{rot_order}
\bibinfo{author}{\bibfnamefont{T.}~\bibnamefont{Cui}},
  \bibinfo{author}{\bibfnamefont{E.}~\bibnamefont{Cheng}},
  \bibinfo{author}{\bibfnamefont{B.}~\bibnamefont{Adler}}, \bibnamefont{and}
  \bibinfo{author}{\bibfnamefont{K.}~\bibnamefont{Whaley}},
  \bibinfo{journal}{Phys. Rev. B} \textbf{\bibinfo{volume}{55}},
  \bibinfo{pages}{12253} (\bibinfo{year}{1997}).

\bibitem[{\citenamefont{Duane et~al.}(1987)\citenamefont{Duane, Kennedy,
  Pendleton, and Roweth}}]{hmc}
\bibinfo{author}{\bibfnamefont{S.}~\bibnamefont{Duane}},
  \bibinfo{author}{\bibfnamefont{A.}~\bibnamefont{Kennedy}},
  \bibinfo{author}{\bibfnamefont{B.}~\bibnamefont{Pendleton}},
  \bibnamefont{and} \bibinfo{author}{\bibfnamefont{D.}~\bibnamefont{Roweth}},
  \bibinfo{journal}{Phys. Lett. B} \textbf{\bibinfo{volume}{195}},
  \bibinfo{pages}{216} (\bibinfo{year}{1987}).

\bibitem[{\citenamefont{Tuckerman et~al.}(1993)\citenamefont{Tuckerman, Berne,
  Martyna, and Klein}}]{hmc2}
\bibinfo{author}{\bibfnamefont{M.}~\bibnamefont{Tuckerman}},
  \bibinfo{author}{\bibfnamefont{B.}~\bibnamefont{Berne}},
  \bibinfo{author}{\bibfnamefont{G.}~\bibnamefont{Martyna}}, \bibnamefont{and}
  \bibinfo{author}{\bibfnamefont{M.}~\bibnamefont{Klein}}, \bibinfo{journal}{J.
  Chem. Phys.} \textbf{\bibinfo{volume}{99}}, \bibinfo{pages}{2796}
  (\bibinfo{year}{1993}).

\bibitem[{\citenamefont{Tuckerman et~al.}(1992)\citenamefont{Tuckerman, Berne,
  and Martyna}}]{multiple-ts}
\bibinfo{author}{\bibfnamefont{M.}~\bibnamefont{Tuckerman}},
  \bibinfo{author}{\bibfnamefont{B.}~\bibnamefont{Berne}}, \bibnamefont{and}
  \bibinfo{author}{\bibfnamefont{G.}~\bibnamefont{Martyna}},
  \bibinfo{journal}{J. Chem. Phys.} \textbf{\bibinfo{volume}{97}},
  \bibinfo{pages}{1990} (\bibinfo{year}{1992}).

\bibitem[{\citenamefont{Matubayasi and Nakahara}(1999)}]{rot-ts}
\bibinfo{author}{\bibfnamefont{N.}~\bibnamefont{Matubayasi}} \bibnamefont{and}
  \bibinfo{author}{\bibfnamefont{M.}~\bibnamefont{Nakahara}},
  \bibinfo{journal}{J. Chem. Phys.} \textbf{\bibinfo{volume}{110}},
  \bibinfo{pages}{3291} (\bibinfo{year}{1999}).

\bibitem[{\citenamefont{Wang et~al.}(1997)\citenamefont{Wang, Johnson, and
  Broughton}}]{pigcmc}
\bibinfo{author}{\bibfnamefont{Q.}~\bibnamefont{Wang}},
  \bibinfo{author}{\bibfnamefont{J.}~\bibnamefont{Johnson}}, \bibnamefont{and}
  \bibinfo{author}{\bibfnamefont{J.}~\bibnamefont{Broughton}},
  \bibinfo{journal}{J. Chem. Phys.} \textbf{\bibinfo{volume}{107}},
  \bibinfo{pages}{5108} (\bibinfo{year}{1997}).

\bibitem[{\citenamefont{Murad and Gubbins}(1978)}]{Murad78}
\bibinfo{author}{\bibfnamefont{S.}~\bibnamefont{Murad}} \bibnamefont{and}
  \bibinfo{author}{\bibfnamefont{K.}~\bibnamefont{Gubbins}}, in
  \emph{\bibinfo{booktitle}{Computer Modelling of Matter}}
  (\bibinfo{publisher}{American Chemical Society},
  \bibinfo{address}{Washington}, \bibinfo{year}{1978}),
  vol.~\bibinfo{volume}{86} of \emph{\bibinfo{series}{ACS Symposium series}},
  p.~\bibinfo{pages}{62}.

\bibitem[{\citenamefont{Steele}(1978)}]{Steele78}
\bibinfo{author}{\bibfnamefont{W.}~\bibnamefont{Steele}}, \bibinfo{journal}{J.
  Phys. Chem.} \textbf{\bibinfo{volume}{82}}, \bibinfo{pages}{817}
  (\bibinfo{year}{1978}).

\end{thebibliography}
\end{document}